\documentclass[aps,twocolumn,superscriptaddress,amsmath,amssymb,showpacs]{revtex4}

\usepackage[english]{babel}
\usepackage{times}
\usepackage{oldgerm,amsmath,pifont,amsfonts,amssymb,latexsym}
\usepackage{float,xspace,calc,theorem,ifthen}
\usepackage{enumerate,psboxit,subfigure}
\usepackage[dvips]{epsfig,graphicx}
\usepackage{hyperref}
\usepackage{changebar}
\usepackage{verbatim}

\newcommand{\rz}{{\mathbb R}}
\newcommand{\nz}{{\mathbb N}}
\newcommand{\zz}{{\mathbb Z}}
\newcommand{\quotemarks}[1]{`\emph{#1}'}

\begin{document}

\title{Capturing correlations in chaotic diffusion by approximation methods}

\author{Georgie Knight}
\email{g.knight@qmul.ac.uk}
\affiliation{School of Mathematical Sciences, Queen Mary University of London,
Mile End Road, London E1 4NS, UK}
\author{Rainer Klages}
\email{r.klages@qmul.ac.uk}
\affiliation{School of Mathematical Sciences, Queen Mary University of London,
Mile End Road, London E1 4NS, UK}

\begin{abstract}

We investigate three different methods for systematically
approximating the diffusion coefficient of a deterministic random walk
on the line which contains dynamical correlations that change
irregularly under parameter variation. Capturing these correlations by
incorporating higher order terms, all schemes converge to the
analytically exact result. Two of these methods are based on expanding
the Taylor-Green-Kubo formula for diffusion, whilst the third method
approximates Markov partitions and transition matrices by using a
slight variation of the escape rate theory of chaotic diffusion. We
check the practicability of the different methods by working them out
analytically and numerically for a simple one-dimensional map, study
their convergence and critically discuss their usefulness in
identifying a possible fractal instability of parameter-dependent
diffusion, in case of dynamics where exact results for the diffusion
coefficient are not available.

\end{abstract}

\maketitle

\section{Introduction}
\label{sec:intro}

Diffusion is a fundamental macroscopic transport process in
many-particle systems. It is quantifiable by the diffusion
coefficient, which describes the linear growth in the mean-squared
displacement of an ensemble of particles. The source of this growth is
often considered to be a Brownian or random process of collisions
between particles. However, on a microscopic scale the equations
governing these collisions in physical systems are deterministic and
typically chaotic. By studying diffusion in chaotic dynamical systems
we can attempt to take these deterministic rules into account and
understand the phenomenon of diffusion from first principles
\cite{Do99,Gasp,Kla06,CAMTV01}. Of particular interest is the study of
the diffusion coefficient under parameter variation in chaotic
dynamical systems such as one-dimensional maps
\cite{GeNi82,SFK,GF2,KoKl03}, area preserving two dimensional maps
\cite{ReWi80,CM81,Ven07} and particle billiards
\cite{MaZw83,HaGa01,HaKlGa02,MaKl03}. Where exact analytical results
for chaotic dynamical systems exist
\cite{RKdiss,RKD,KlDo99,GrKl02,Crist06,Kni11}, one finds that the
diffusion coefficient is typically a complicated fractal function of
control parameters. This phenomenon can be understood as a topological
instability of the deterministic diffusive dynamics under parameter
variation \cite{RKdiss,RKD,KlDo99,Kla06}.

So far exact analytical solutions for the diffusion coefficient could
only be derived for simple cases of low-dimensional dynamics.  In
higher dimensions even very fundamental properties of diffusion
coefficients are often unknown, such as whether they are smooth or
fractal functions of control parameters
\cite{HaGa01,JeRo06,Kla06}. For example, much effort was spent two
decades ago studying more complicated systems like a two-dimensional
family of sawtooth maps \cite{CM81,DMP89,Eck92}. However, despite a
good understanding of the orbit structure \cite{PV86,PV862} it was not
possible to conclude whether the diffusion coefficient is fractal or
not \cite{Sano02}. If one wishes to achieve a microscopic
understanding of diffusion in more realistic physical systems, one
therefore has to rely either on numerical simulations or on
approximation methods.

In this paper we compare three different methods for approximating
parameter dependent diffusion coefficients with each other by working
them out analytically and numerically for a simple one-dimensional
map. This model has the big advantage that it is very amenable to
rigorous analysis. Its diffusion coefficient has been calculated
exactly in \cite{Kni11} and was found to be a fractal function of a
control parameter. Our goal is to assess the individual capabilities
and limitations of these approximation methods in terms of
practicability, physical interpretation, convergence towards the exact
result, and identification of an underlying fractal structure in the
diffusion coefficient. We also address recent criticism by Gilbert and
Sanders \cite{Gil09}, who claimed that one of these methods, as
originally proposed in \cite{KlKo02}, is mathematically wrong and
unphysical.

In section \ref{sec:sys} we define the deterministic dynamical system
that provides our test case, which is a simple piecewise linear
one-dimensional map.  In section \ref{sec:trunc} the first
approximation method is introduced, called correlated random walk in
\cite{KlKo02}, which consists of truncating the Taylor-Green-Kubo
formula for diffusion. This method enables us to analytically build up
a series of approximations which gives evidence for a fractal
structure.  In previous work this approximation scheme has
successfully been applied to understand parameter dependent diffusion
in models that are much more complicated than the one considered here
\cite{MaKl03,KoKl03,HaKlGa02,Kla06}. Motivated by the criticism of
\cite{Gil09}, in this paper we provide further insight into the
functioning of this method by working it out rigorously for our
specific example. In section \ref{sec:pers} the persistent random walk
method for diffusion is studied. This method was originally proposed
within stochastic theory in the form of a persistent random walk
\cite{HK87,Weiss94}. It consists of approximating the
Taylor-Green-Kubo formula by including memory in a self-consistent,
persistent way. Recently this method has been worked out for chaotic
diffusion in Hamiltonian particle billiards
\cite{Gil09,Gil10,Gil11}. Here we apply this scheme to the different
case of a one-dimensional map, and we obtain a series of
approximations analytically and then numerically. In section
\ref{sec:approx_mark} we look at a third method, defined by a
slight variation \cite{RKD,RKdiss,KlDo99} of the escape rate theory of
chaotic diffusion \cite{GN,GaDo95,BTV96,Gasp,Do99,Voll02} in that
absorbing boundary conditions are replaced by periodic ones. This
method thus consists of evaluating the diffusion coefficient in terms
of the decay rate of the dynamical system towards the equilibrium
state, instead of using the escape rate. The decay rate is in turn
obtained by an approximation to the relevant Markov transition
matrix. By this method we are able to build up a series of
approximations which, through the functional form of the interpolation
that we find, gives very strong evidence for fractality. Basic ideas
defining this method have been sketched in \cite{Kla06}, however, this
is the first time that it has been fully worked out to understand
fractal diffusion coefficients. Section \ref{sec:conclusion} forms the
conclusion.

\section{A one-dimensional map exhibiting chaotic diffusion}
\label{sec:sys}

\begin{figure*}[htb]
\begin{center}
  \includegraphics[height=7cm]{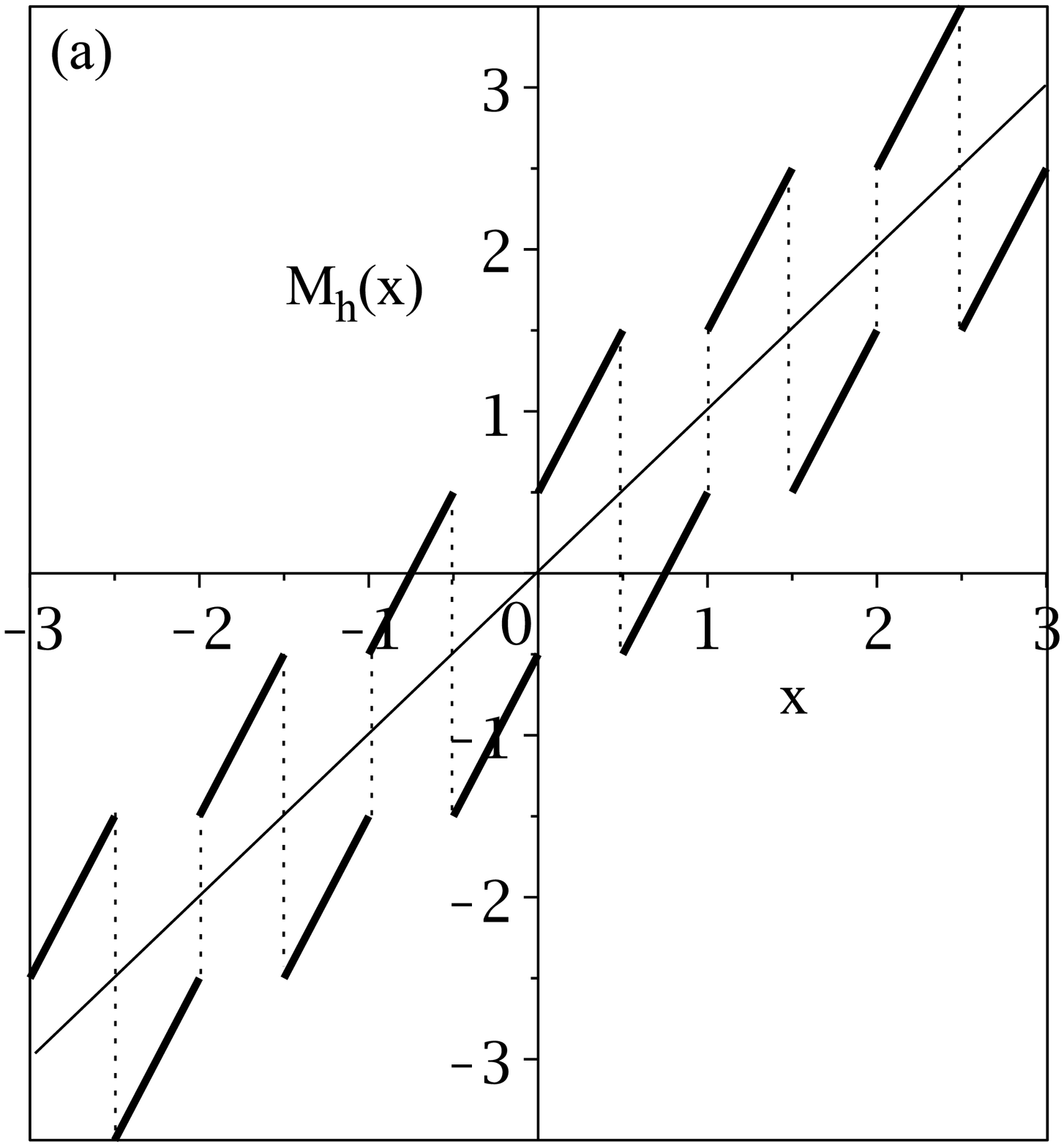}\includegraphics[height=7cm]{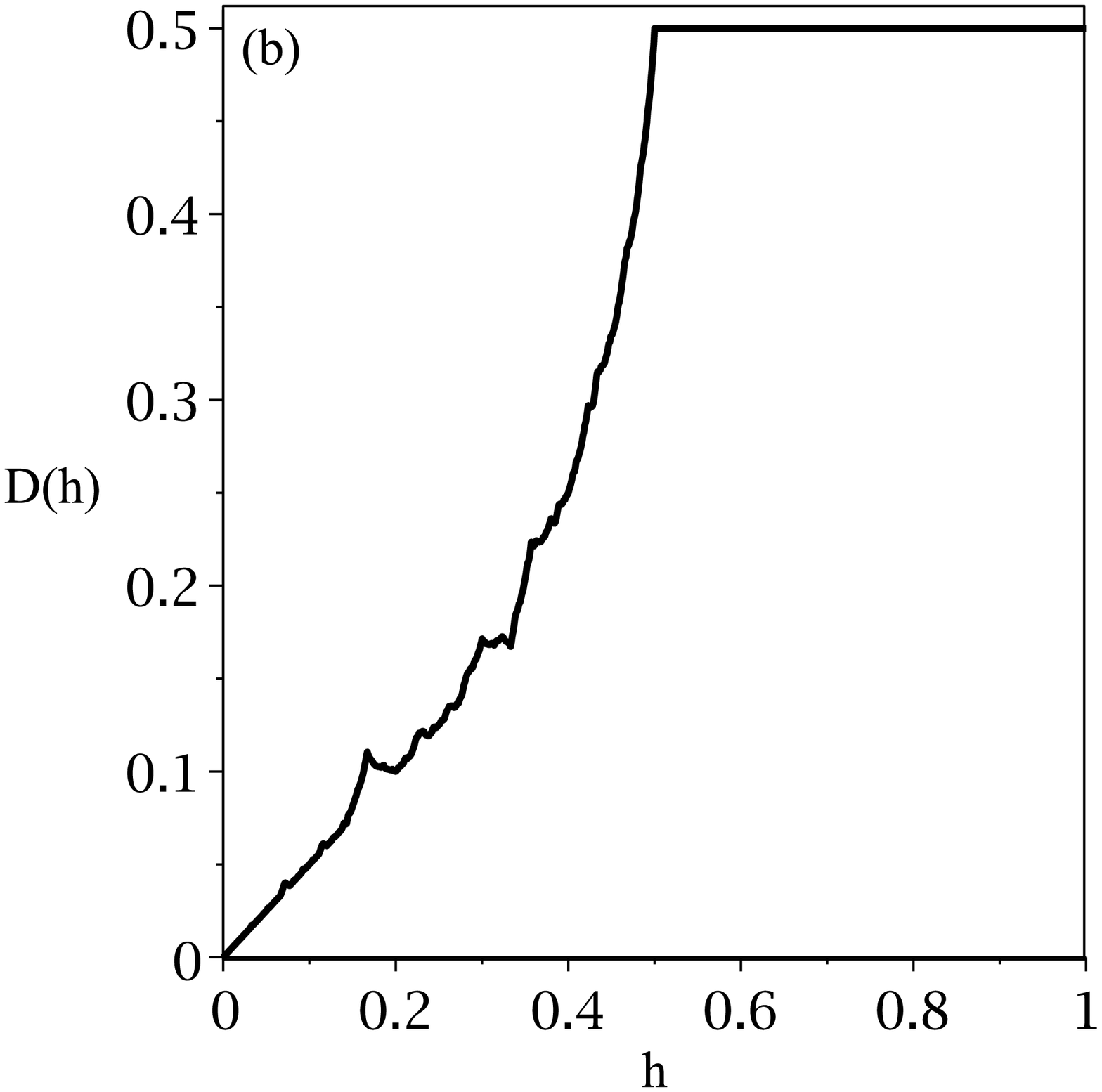}
\end{center}
\caption{The lifted Bernoulli shift map. A section of the
map $M_h(x)$, Eqs.~(\ref{Eq:M_h_box}) and (\ref{Eq:lift}),
is illustrated in \textbf{(a)} for the
value of the control parameter $h=0.5$. The corresponding parameter
dependent diffusion coefficient $D(h)$, exactly calculated in
\cite{Kni11}, is shown in \textbf{(b)}.}
\label{Fig:map_diff}
\end{figure*}

We use the simplest setting possible, where deterministic diffusion is
generated by a parameter dependent one-dimensional dynamical
system. The equations of motion are determined by a map $M_h(x):\rz
\rightarrow \rz$ so that
\begin{eqnarray}\nonumber
                x_{n+1}&=&M_h(x_n)\\
                       &=& M^{n+1}_h(x) \ \ x\in\rz, h\ge0, n\in \nz,
\label{Eq:x_iter}
\end{eqnarray}
with $x=x_0$ \cite{GeNi82,SFK,GF2,RKD}. In our case, the map $M_h(x)$
is based on the {\em Bernoulli shift} or {\em doubling
map}, combined with a lift parameter $h$, which gives the
simple parameter dependent map of the interval
\begin{equation}
M_h (x)=
\left\{
\begin{array}{rl}
2x + h & 0\leq x <\frac{1}{2}\\
2x -1 -h & \frac{1}{2}\leq x < 1\end{array}\right. .
\label{Eq:M_h_box}
\end{equation}
This map exhibits `escape', i.e., points leave the unit interval under
iteration. It is copied and lifted over the real line by
\begin{equation}
                M_h(x+z)=M_h(x)+z, \ \ z\in \zz
\label{Eq:lift}
\end{equation}
in order to obtain a map from the real line to itself, see
Fig.~\ref{Fig:map_diff}(a). The symmetry in this system ensures that
there is no mean drift \cite{GrKl02}. Note that the invariant density
of the map Eq.~(\ref{Eq:M_h_box}) modulo 1 remains by construction
simply uniform throughout the whole parameter range. This is in
contrast to the related piecewise linear maps studied in
\cite{RKdiss,RKD,KlDo99,Kla06}, where the density becomes a highly
complicated step function under parameter variation, which profoundly
simplifies the situation. The model was first introduced in
\cite{GaKl}, where its parameter dependent diffusion coefficient
$D(h)$ was obtained numerically, while in \cite{PG1,Do99} the
diffusion coefficient for a special single parameter value was
calculated analytically.  Exact analytical solutions for $D(h)$ for
all $h\ge0$ of this and related models were recently obtained in
\cite{Kni11}. Since there is a periodicity with integer values of $h$,
here we restrict ourselves to the parameter regime of $h \in [0,1]$
without loss of generality. In \cite{GaKl,Kni11} it was found that
$D(h)$ displays both fractal and linear behaviour, see
Fig.~\ref{Fig:map_diff}(b). To our knowledge, this is one of the
simplest models that exhibits a fractal diffusion coefficient. Being
nevertheless amenable to rigorous analysis, it thus forms a convenient
starting point to learn about the power of different approximation
methods for understanding complicated diffusion coefficients.

\section{Correlated random walk}
\label{sec:trunc}

\begin{figure*}[htb]
\begin{center}
  \includegraphics[height=7cm]{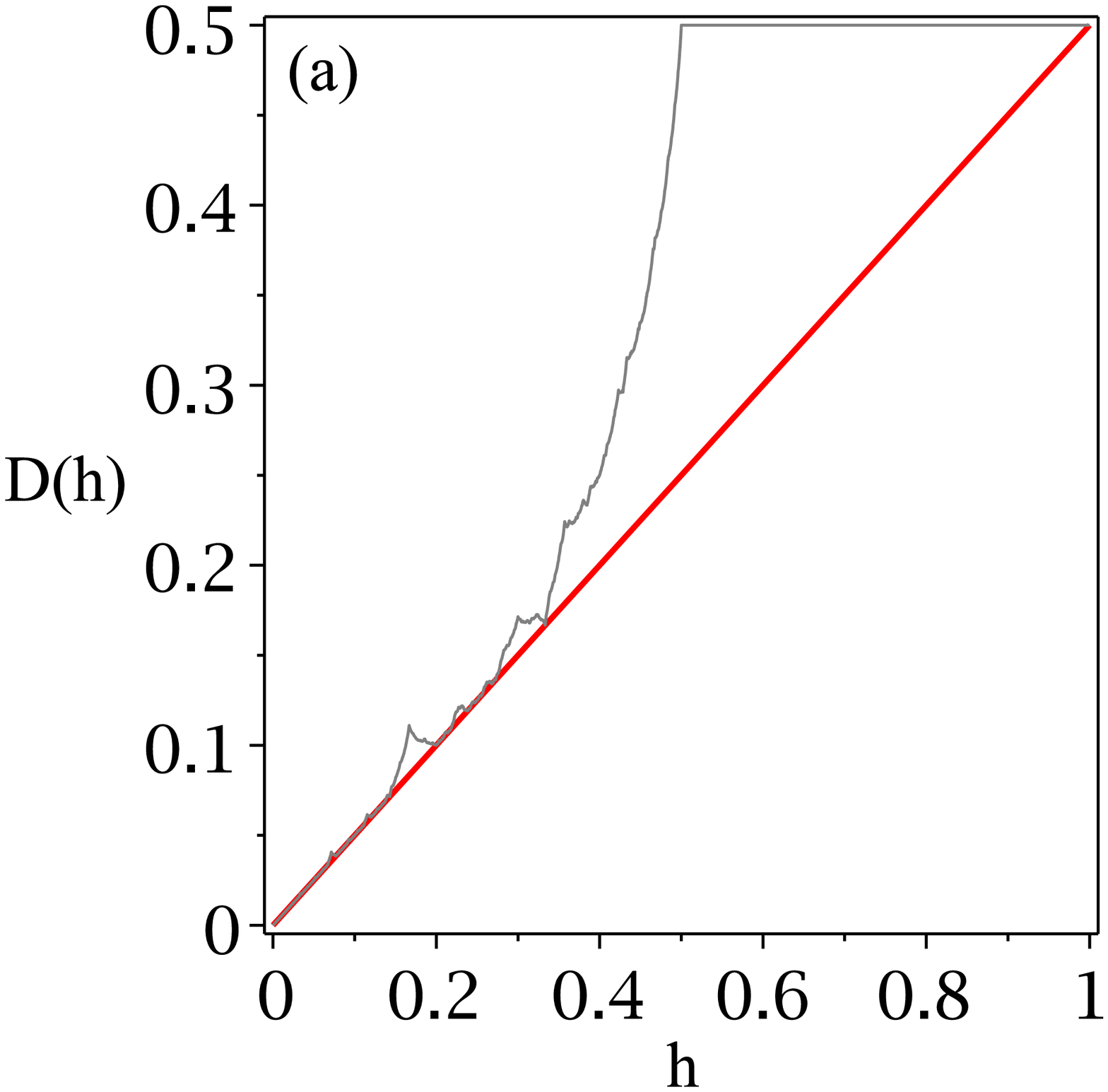} \includegraphics[height=7cm]{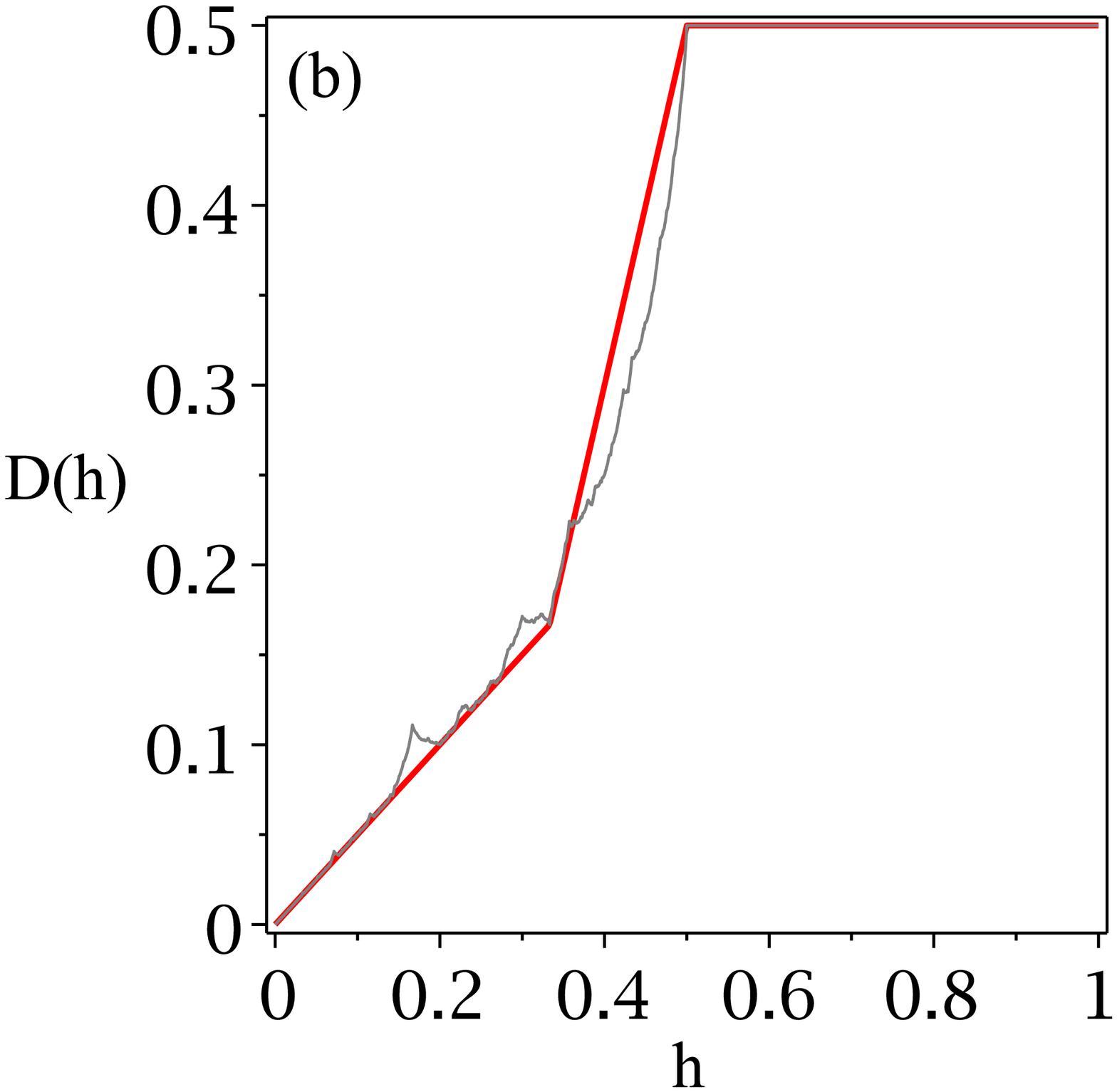}\\
  \includegraphics[height=7cm]{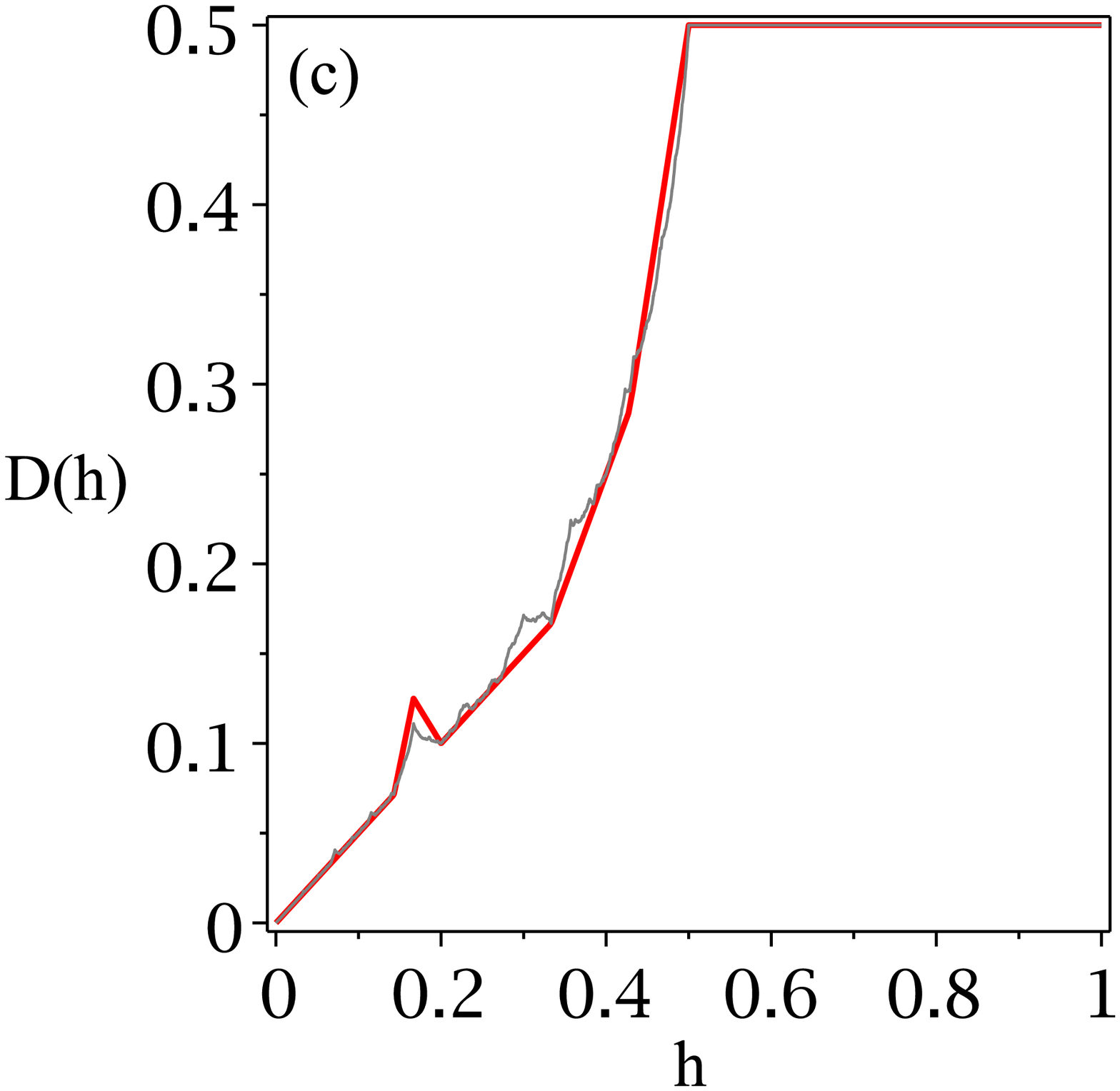} \includegraphics[height=7cm]{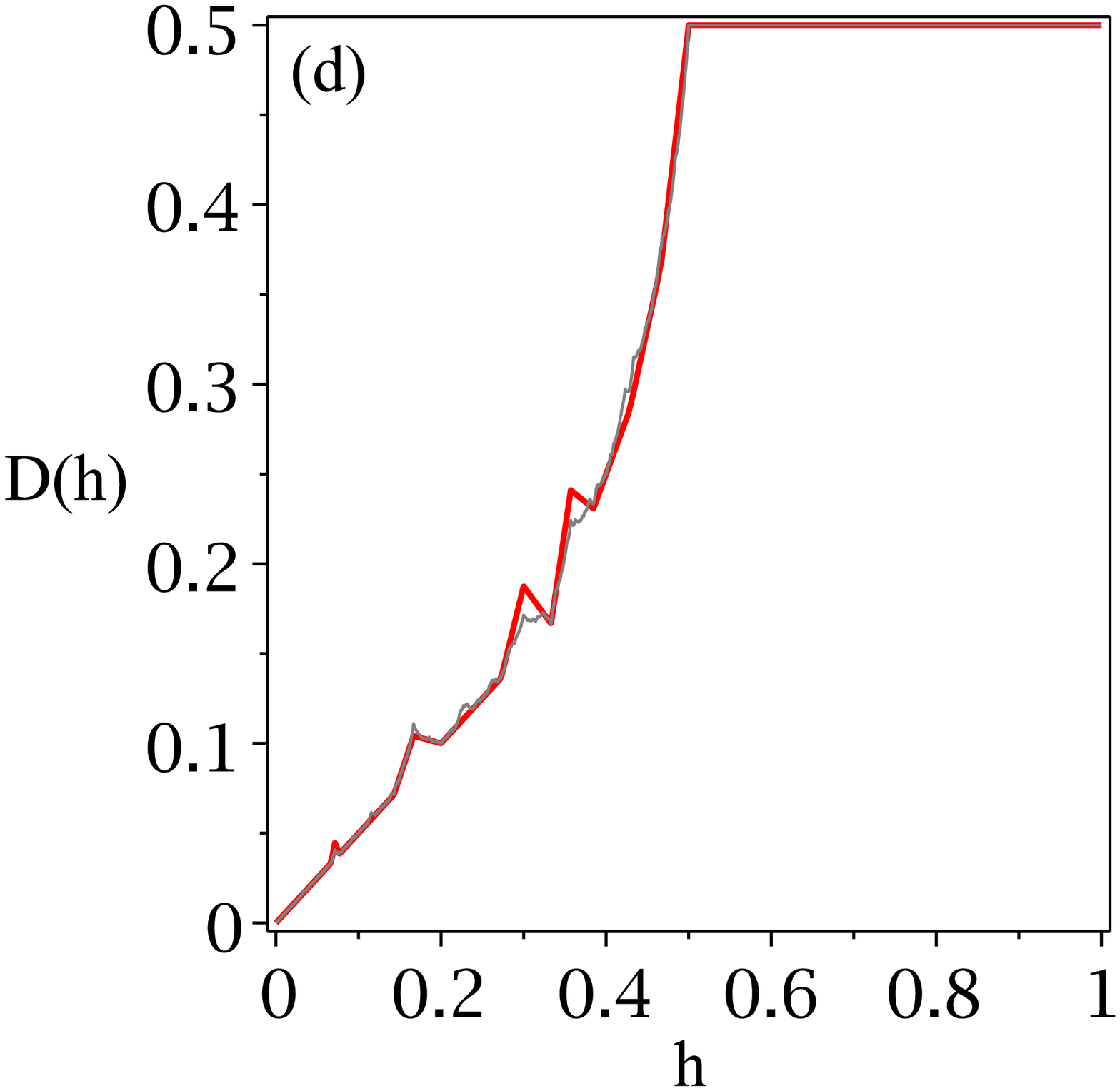}
\end{center}
\caption{(Colour online) Correlated random walk .  In this figure the
first four approximations to the parameter dependent diffusion
coefficient $D(h)$ are illustrated in bold (red) along with the actual diffusion coefficient. In \textbf{(a)} the zeroth order
is shown, which is simply the random walk solution, in
\textbf{(b)},\textbf{(c)} and \textbf{(d)} the first, second, and
third order approximations, respectively. At each stage one obtains a
set of extrema with linear interpolation, which converge quickly
to the exact diffusion coefficient $D(h)$. The amount of extrema
increases exponentially with $n$, hence we see the fractal structure
emerging.}
\label{Fig:Trunc_TGK}
\end{figure*}

Our first approximation method starts with the diffusion coefficient
expressed in terms of the velocity autocorrelation function of the
map, called the Taylor-Green-Kubo formula, see \cite{Do99,KlKo02} for
derivations,
\begin{eqnarray}\nonumber
               D(h) &=& \lim_{n\to\infty} \left(\sum_{k=0}^n \int_0^1 v_0(x) v_k(x) \rho^*(x) dx \right) \\
 & &-\frac{1}{2}\int_0^1 v_0^2(x) \rho^*(x) dx\:,
\label{Eq:TGK}
\end{eqnarray}
where $\rho^*(x)$ is the invariant density of the map
Eq.~(\ref{Eq:M_h_box}) modulo 1, this being equal to one throughout the
parameter range as we have a family of doubling maps. The velocity
function $v_k(x)$ calculates the integer displacement of a point at
the $k^{th}$ iteration,
\begin{equation}
               v_k(x)=\left\lfloor x_{k+1}\right\rfloor- \left\lfloor x_{k}\right\rfloor.
\label{Eq:v_x}
\end{equation}
In order to create an $n^{th}$ order approximation we simply truncate
Eq.~(\ref{Eq:TGK}) at a given $n$ \cite{KlKo02}. Hence we obtain the finite sum
\begin{equation}
                D_n(h)=\sum_{k=0}^n \int_0^1 v_0(x)v_k(x)dx - \frac{1}{2}\int_0^1 v^2_0(x)dx\:,
\label{Eq:TGK_trunc}
\end{equation}
which can physically be understood as a time dependent diffusion
coefficient.  Looking at how the sequence of $D_n(h)$ converges
towards $D(h)$ thus corresponds to incorporating more and more memory
in the decay of the velocity autocorrelation function and checking how
this decay varies as a function of $h$ for a given $n$.  Note that the
functional form of $D_n(h)$ for finite $n$ is to some extent already
determined by our choice of integer displacements in
Eq.~(\ref{Eq:v_x}), however, we have checked that for the given model
the deviations between using integer and non-integer displacements for
finite time are minor. Secondly, we remark that by using this
straightforward truncation scheme we have neglected further
cross-correlation terms that do not grow linearly in $n$, cf.\
\cite{Do99}. Still, by definition we have $D_n(h)\to
D(h)\:(n\to\infty)$. Going to lowest order, for $n=0$ we immediately
see that
\begin{equation}
               D_0(h)=\frac{h}{2}\:,
\label{Eq:d0h}
\end{equation}
which is the simple uncorrelated random walk solution for the
diffusion coefficient \cite{Kni11}. In Fig.~(\ref{Fig:Trunc_TGK}) one
can see that $D_0(h)$ is asymptotically exact for $h\rightarrow 0$.

Of more interest however are the higher values of $n$ capturing the
higher order correlations that come into play. To evaluate these we
define a jump function $J^n_h(x):[0,1]\rightarrow \rz$,
\begin{equation}
                J^n_h(x)=\sum_{k=0}^{n} v_k\left(x\right)\:,
\label{Eq:jump}
\end{equation}
which gives the integer displacement of a point $x$ after $n$
iterations. Equation (\ref{Eq:jump}) can be written recursively as
\cite{Kni11}
\begin{equation}
                J^n_h(x)= v_0(x)+ J^{n-1}_h\left(\tilde{M}_h(x)\right),
\label{Eq:jump_rec}
\end{equation}
where $\tilde{M}_h(x)$ is Eq.~(\ref{Eq:M_h_box}) taken modulo
$1$. This recursive formula will help when we solve the integral in
Eq.~(\ref{Eq:TGK_trunc}). Let $T_h^n(x):[0,1]\rightarrow\rz$ be
defined as
\begin{equation}
                T^n_h(x)= \int_0^x J^n(y)dy, \ \ T^{-1}_h(x):=0.
\label{Eq:tak_def}
\end{equation}
Using Eq.~(\ref{Eq:jump_rec}) we can solve
Eq.~(\ref{Eq:tak_def}) recursively as
\begin{equation}
                T^n_h(x)= s_h(x)+ \frac{1}{2}T^{n-1}\left(\tilde{M}_h(x)\right)
\label{Eq:tak_def_recurs}
\end{equation}
with
\begin{equation}
                s_h(x)=\int_0^x v_0(y)dy= xv_0(x)+c,
\label{Eq:s(x)}
\end{equation}
where the constants of integration $c$ can be evaluated using the
continuity of $T_h^n(x)$ and the fact that $T_h^n(0)=T_h^n(1)=0$ as
there is no mean drift in this system. We obtain the following
functional recursion relation for $T^n_h(x)$:
\begin{widetext}
\begin{equation}
T_h^n (x)=
\left\{
\begin{array}{lll}
\frac{1}{2}T^{n-1}_h(2x+h)  &-\frac{1}{2}T_h^{n-1}(h)                                 &\ \ 0 \leq x < \frac{1-h}{2}\\
\frac{1}{2}T^{n-1}_h(2x+h-1)&-\frac{1}{2}T_h^{n-1}(h)+x +\left(\frac{h-1}{2}\right)   &\ \ \frac{1-h}{2} \leq x < \frac{1}{2}\\
\frac{1}{2}T^{n-1}_h(2x-h)  &-\frac{1}{2}T_h^{n-1}(h)-x +\left(\frac{h+1}{2}\right)   &\ \ \frac{1}{2}\leq x < \frac{1+h}{2}\\
\frac{1}{2}T^{n-1}_h(2x-1-h)&-\frac{1}{2}T_h^{n-1}(h)                                 &\ \ \frac{1+h}{2} \leq x  < 1 \end{array}\right. \ \ .
\label{Eq:tak_full}
\end{equation}
\end{widetext}
Using Eq.~(\ref{Eq:tak_full}) in Eq.~(\ref{Eq:TGK_trunc})
via Eqs.~(\ref{Eq:jump}) and (\ref{Eq:tak_def}) we can evaluate
our $n^{th}$ order approximation as
\begin{equation}
                D_n(h)=  \frac{h}{2} + T_h^{n-1}(h).
\label{Eq:D_n(h)}
\end{equation}
So we see that the higher order correlations are all captured by the
cumulative integral functions $T^n_h(x)$. In order to evaluate
Eq.~(\ref{Eq:D_n(h)}) we construct a recursive relation from
Eq.~(\ref{Eq:tak_full}),
\begin{equation}
                T^n_h(h) = \sum_{k=0}^{n} \frac{1}{2^k}t_h\left(\tilde{M}_h^k(h)\right)- \sum_{k=1}^{n}\frac{1}{2^k}T_h^{n-k}(h),
\label{Eq:tak_recurs}
\end{equation}
where
\begin{equation}
t_h (x)=
\left\{
\begin{array}{rl}
0                    &\ \ 0 \leq x < \frac{1-h}{2}\\
x  +\frac{h-1}{2}    &\ \ \frac{1-h}{2} \leq x < \frac{1}{2}\\
-x +\frac{h+1}{2}                 &\ \ \frac{1}{2}\leq x < \frac{1+h}{2}\\
0                    &\ \ \frac{1+h}{2} \leq x < 1 \end{array}\right.
\label{Eq:t_h(x)}
\end{equation}
is $s_h(x)$ with the $-\frac{1}{2}T_h^n(h)$ terms removed. In order to
simplify Eq.~(\ref{Eq:tak_recurs}) we write it entirely in terms
of Eq.~(\ref{Eq:t_h(x)}). Let
\begin{equation}
                \tau_h(n)=\sum_{k=1}^{n}\frac{1}{2^k}T_h^{n-k}(h).
\label{Eq:tau(n)}
\end{equation}
We can write Eq.~(\ref{Eq:tau(n)}) recursively as
\begin{equation}
                \tau_h(n)= \frac{1}{2}T_h^{n-1}(h)+\frac{1}{2}\tau_h(n-1).
\label{Eq:tau(n)_recurs}
\end{equation}
Substituting Eqs.~(\ref{Eq:tau(n)}) and (\ref{Eq:tau(n)_recurs})
into (\ref{Eq:tak_recurs}) we obtain
\begin{equation}
                T^n_h(h) = \sum_{k=0}^{n} \frac{1}{2^k}t_h\left(\tilde{M}_h^k(h)\right)- \frac{1}{2}T_h^{n-1}(h)-\frac{1}{2}\tau_h(n-1).
\label{Eq:tak_rec_2}
\end{equation}
Then substituting Eq.~(\ref{Eq:tak_recurs}) back into Eq.~(\ref{Eq:tak_rec_2})

\begin{widetext}
\begin{eqnarray}
                T^n_h(h) = \sum_{k=0}^{n} \frac{1}{2^k}t_h\left(\tilde{M}_h^k(h)\right)- \frac{1}{2}\left(  \sum_{k=0}^{n-1} \frac{1}{2^k}t_h\left(\tilde{M}_h^k(h) \right)\right)+\frac{1}{2}\tau_h(n-1) -\frac{1}{2}\tau_h(n-1)
\label{Eq:tak_rec_3}
\end{eqnarray}
\end{widetext}

we arrive at our final expression
\begin{equation}
               T^n_h(h)=\frac{1}{2^{n}}t_h\left(\tilde{M}_h^n(h)\right)+\sum_{k=0}^{n-1} \frac{1}{2^{k+1}}t_h\left(\tilde{M}_h^k(h)\right).
\label{Eq:tak_recurs_final}
\end{equation}

It is helpful to rewrite Eq.~(\ref{Eq:tak_recurs}) in the form of
Eq.~(\ref{Eq:tak_recurs_final}) as it allows us to show that under
this method, $D_n(h)$ converges exactly to $D(h)$ in finite time for
particular values of $h$, see Fig.~(\ref{Fig:Trunc_TGK}) for an
illustration. This means that for a specific set of parameter values,
we can fully capture the correlations of the map with a finite time
approximation. This convergence is dependent upon the behaviour of the
orbit of the point $x=h$ under the map $\tilde{M}_h(x)$.  In
particular, if this orbit is pre-periodic, and the values of the
points in the periodic loop correspond to $0$ in
Eq.~(\ref{Eq:t_h(x)}), then the time dependent diffusion coefficient
$D_n(h)$ will converge to the exact value $D(h)$ on the $n^{th}$ step,
where $n$ is given by the transient length of the orbit of $h$ plus
one. For example, let $h=2/5$,
\begin{eqnarray}\nonumber
               \tilde{M}_{2/5}(2/5)    &=& 1/5\\\nonumber
               \tilde{M}_{2/5}(1/5)    &=& 4/5\\
               \tilde{M}_{2/5}(4/5)    &=& 1/5\:.
\label{Eq:orbit_2/5}
\end{eqnarray}
So $h=2/5$ is pre-periodic of transient length one. In addition
\begin{eqnarray}\nonumber
               t_h(2/5)    &=& 1/10\\ \nonumber
               t_h(1/5)    &=& 0\\
               t_h(4/5)    &=& 0\:,
\label{Eq:t_1/3}
\end{eqnarray}
thus $t_{2/5}\left(\tilde{M}_{2/5}^n(2/5)\right)=0$ for $n>1$. Hence
we see finite time convergence to $D(h)$. This finite time convergence
at a certain set of points is helpful in understanding the structure of $D(h)$
as the fractal diffusion coefficient can be seen emerging around these
points in the same manner as an iterated function system like a Koch
curve, see Fig.~(\ref{Fig:Trunc_TGK}).

Being able to analytically expose the fractal structure of parameter
dependent diffusion coefficients is the main strength of this
method. In addition, the convergence of the series of approximations
is very quick due to the finite time convergence at certain values of
$h$. Moreover, the fact that one only needs to directly put in the map
dynamics makes it very user-friendly. However, due to the recurrence
relation that this method is based on, applying it analytically is
restricted to one-dimensional systems or higher dimensional systems
whose dynamics can be projected down to one-dimensional systems, such
as baker maps \cite{GaKl,Gasp,Do99,Voll02,Kla06}. In order to answer
questions about more realistic, physical systems, one would need to
resort to numerical analysis. By using families of time and parameter
dependent diffusion coefficients such as defined by
Eq.~(\ref{Eq:TGK_trunc}) this is, on the other hand, straightforward,
as has been successfully demonstrated for many different types of
systems \cite{MaKl03,KoKl03,HaKlGa02,KlKo02,Kla06}.

This approximation method, represented by Eq.~(\ref{Eq:TGK_trunc}),
was criticized by Gilbert and Sanders in \cite{Gil09} in two ways:
First, it was stated that \quotemarks{this ad hoc truncation has no
physical meaning: if $<v_0v_l>\neq0$, it is not true that higher-order
correlations $<v_0v_k>$ vanish}. However, there is no assumption in
Eq.~(\ref{Eq:TGK_trunc}) that higher-order correlations disappear. On
the contrary, this expansion is to be truncated at different time
steps for exploring the impact of higher-order correlations on
the convergence of the series by systematically incorporating them
step by step. Interestingly, as we have shown above, there do exist
parameter values for this model at which all higher order correlations
disappear. This set, whose number of elements becomes infinite for
$n\to\infty$, holds the key to understanding the emergence of the fractal
structure in the diffusion coefficient. There is a clear physical
interpretation of this set of parameter values in terms of the orbits
of the associated critical points of the map, as exemplified
above. Under parameter variation these orbits generate complicated
sequences of forward and backward scattering, which characterize the
diffusive dynamics by physically explaining the origin of the fractal
structure in terms of the topological instability of the associated
microscopic scattering processes. This physical interpretation, called
\quotemarks{turnstile dynamics}, has been explained in detail in
\cite{RKdiss,RKD,KlDo99,Kla06,Kni11}.

Secondly, by applying a higher-dimensional equivalent of
Eq.~(\ref{Eq:TGK_trunc}) to the billiard models discussed in
\cite{Gil09} Gilbert and Sanders claimed that in \cite{KlKo02}
\quotemarks{the stationary distribution was erroneously assumed to be
uniform}. We first clarify that there is no room for `assuming' any
stationary distribution in this equation. The mathematically exact
derivation of the Taylor-Green-Kubo formula Eq.~(\ref{Eq:TGK}) is
based on time translational invariance of the dynamics, cf.\
\cite{Do99,KlKo02}, and can only be carried out if the density
$\rho^*(x)$ in Eq.~(\ref{Eq:TGK_trunc}) is strictly the invariant
one. Hence, there is no choice, and for our map as well as for the
Hamiltonian particle billiards studied in \cite{Gil09} this density
has to be the invariant one, which in turn for all these systems is
uniform. Gilbert and Sanders claim to `correct this mistake' by
deriving a second-order approximation of their billiard models which
is different from the one obtained from the method outlined in this
section as applied to billiards, compare Eq.~(11) in \cite{Gil09} with
Eq.~(21) in \cite{KlKo02}. Their Eq.~(11), which they use for their
simulations, thus seems to represent a mix between the method outlined
in this section and the one described in the following section.

\section{Persistent random walk}
\label{sec:pers}

The next method we look at again starts with the Taylor-Green-Kubo
formula for diffusion Eq.~(\ref{Eq:TGK}). However, rather than
truncating it, we now approximate the correlations in a more
self-consistent way by including memory effects persistently. The key
difference to the previous method is that this approach models an
exponential decay of the velocity autocorrelation function beyond the
lowest order approximation. This method first emerged within
stochastic theory as a persistent random walk \cite{HK87,Weiss94} and
was recently applied to understand chaotic diffusion in Hamiltonian
particle billiards \cite{Gil09,Gil10,Gil11}.

The main task of evaluating the diffusion coefficient with this method
is to find an expression for the correlation function at the
$n^{th}$ time step by only including memory effects of a given
length. We start by defining the velocity autocorrelation function as
a sum over all possible velocities weighted by the corresponding parts
of the invariant measure $\mu^*$ of the system,
\begin{equation}
                \left\langle   v_0(x)v_n(x) \right\rangle = \hspace*{-0.2cm}\sum_{v_0(x), \ldots , v_n(x)}\hspace*{-0.7cm} v_0(x)v_n(x) \mu^* (\{v_0(x), \ldots, v_n(x)  \}).
\label{Eq:vel_aut_GS}
\end{equation}
The different parts of the invariant measure in
Eq.~(\ref{Eq:vel_aut_GS}) are approximated by the transition
probabilities of the system, depending on the length of memory
considered. These in turn are trivially obtained from the invariant
probability density function $\rho^*(x)$. As a $0^{th}$ order approximation of
this method, no memory is considered at all, that is, the movement of
a particle is entirely independent of its preceding behaviour. In this
case the correlations evaluate simply as
\begin{equation}
                \left\langle   v_0(x)v_n(x) \right\rangle = 0   \ \ (n>0),
\label{Eq:vel_0_order}
\end{equation}
thus we need only consider $ \left< v^2_0(x)\right>$. By
Eq.~(\ref{Eq:TGK}) the approximate diffusion coefficient is obtained as
\begin{equation}
                D_0(h) = \frac{1}{2}\int_0^1 \left\langle   v^2_0(x) \right\rangle dx
                       = \frac{h}{2}\:,
\label{Eq:GS_Ran_w}
\end{equation}
which reproduces again the random walk solution, as expected. For the
higher order approximations, one must refine the level of memory that
is used based upon the microscopic dynamics of the map.

\subsection{One step memory approximation}
\label{subsec:onestep}

\begin{figure*}[ht]
\begin{center}
\includegraphics[height=7.5cm]{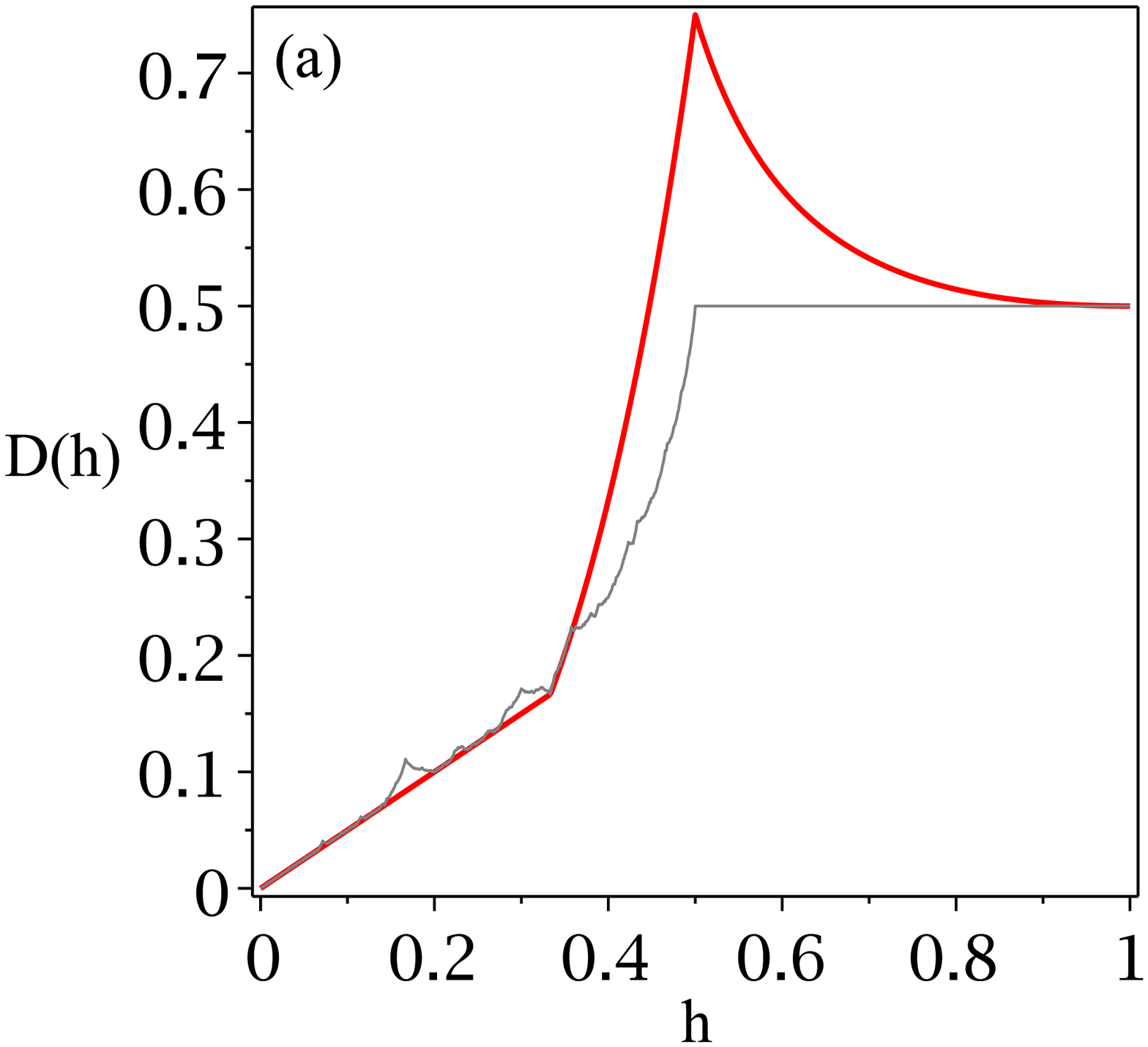} \includegraphics[height=7.5cm]{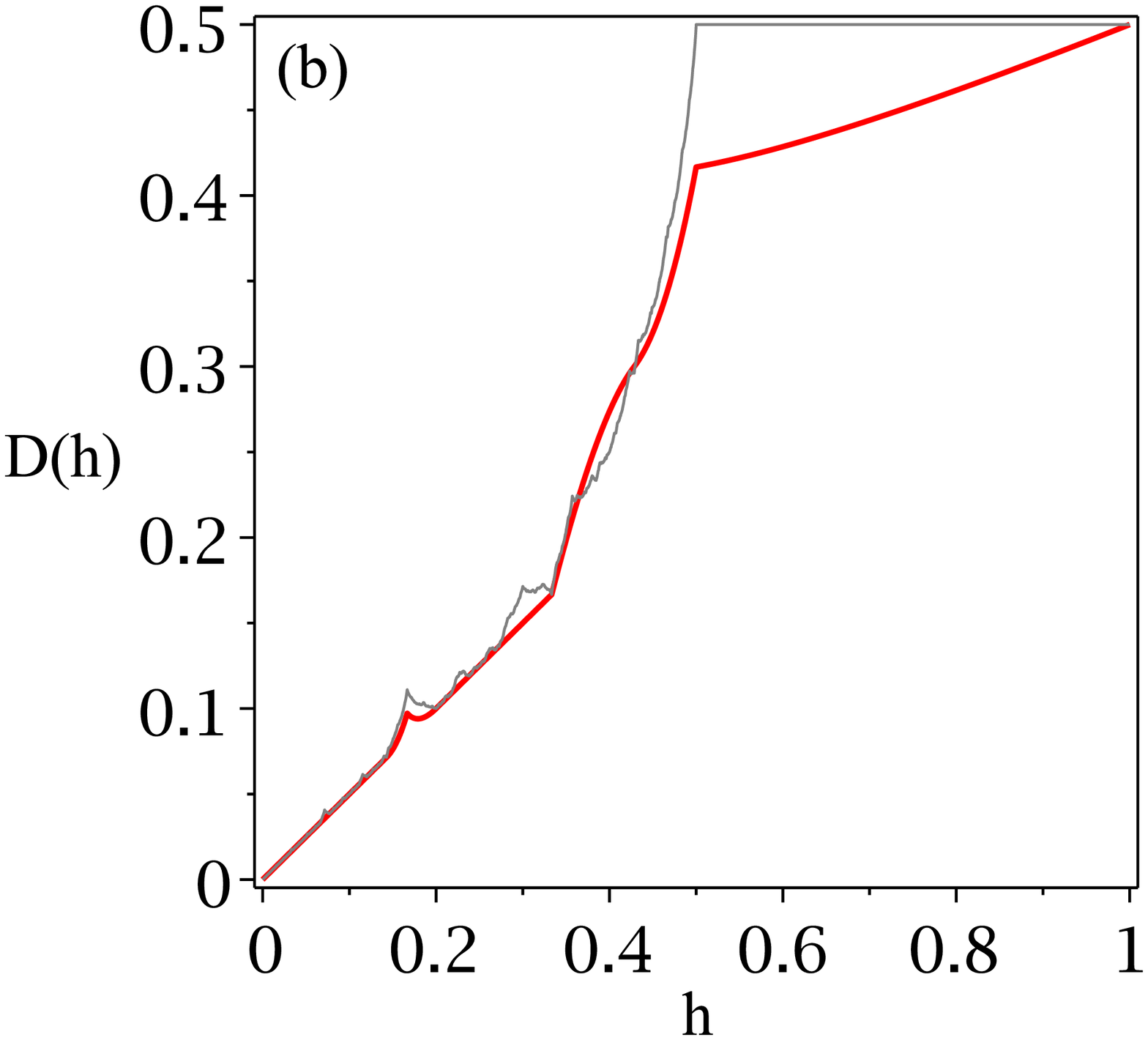}
\end{center}
\caption{(Colour online) Persistent random walk approximation . In this figure the
first order approximation Eq.~(\ref{Eq:1_step_D}) to the exact
parameter dependent diffusion coefficient $D(h)$ is illustrated in
\textbf{(a)}, the second order Eq.~(\ref{Eq:2s}) is shown in
\textbf{(b)}. Approximations are in bold (red) along with the diffusion coefficient. The major topological changes in the dynamics are picked
out by piecewise-differentiable approximations.}
\label{Fig:Pers}
\end{figure*}

We now include one step of memory in the system, i.e., we assume that
the behaviour of a point at the $n^{th}$ step is only dependent on the
$(n-1)^{th}$ step. In \cite{Gil09} Eq.~(\ref{Eq:vel_aut_GS}) was
evaluated for approximating the diffusion coefficient in a particle
billiard. For this purpose it was assumed that a point moves to a
neighbouring lattice point at each iteration. Hence the velocity
function $v_n(x)$ could only take the values $\ell$ or $-\ell$, where
$\ell$ defines the lattice spacing. In order to evaluate the one step
memory approximation for the map $M_h(x)$, we need to modify the
method to include the probability that a point stays at a lattice
point and does not move, hence our velocity function can take the
values $1,-1$ or $0$. Let $P(b|a)$ be the conditional probability that
a point takes the velocity $b$ given that at the previous step it had
velocity $a$ with $a,b \in \left\{0,1,-1\right\}$. We use these
probabilities to obtain a one step memory approximation. We can write
our velocity autocorrelation function as
\begin{equation}
                \left\langle   v_0 v_n \right\rangle = \sum_{v_0, ... , v_n} v_0 v_n  p(v_0) \prod_{i=1}^n P(v_i |v_{i-1}),
\label{Eq:vel_aut_1step}
\end{equation}
where we let $v_k(x)=v_k$ for brevities sake and $p(a)$ is
the probability that a point takes the velocity $a$ at the first
step. We can capture the combinatorics of the sum over all possible
paths by rewriting Eq.~(\ref{Eq:vel_aut_1step}) as a matrix equation,
\begin{equation}
                \left\langle   v_0  v_n \right\rangle =
                 \left(\begin{array}{ccc}
                 0&1&-1\end{array}\right)\hspace*{-0.1cm}
                 \left(\begin{array}{ccc}
                 P_{00} &  P_{01} &  P_{0-1} \\
                 P_{10} &  P_{11} &  P_{1-1} \\
                 P_{-10} &  P_{-11} &  P_{-1-1} \end{array}\right)^n\hspace*{-0.2cm}
                 \left(\begin{array}{c}
                 0 \\
                 p(1) \\
                 -p(-1) \end{array}\right)
\label{Eq:vel_aut_matrix}
\end{equation}

where $P_{ba}=P(b|a)$. Equation (\ref{Eq:vel_aut_matrix}) can be
simplified by using the fact that all the paths with a \quotemarks{0}
state cancel each other out, therefore not contributing to diffusion,
and by using the symmetries in the system, i.e.,
\begin{eqnarray}\nonumber
               P_{-1-1} &=& P_{11}  \\ \nonumber
                P_{-11} &=& P_{1-1} \\
                p(-1)    &=& p(1)
\label{Eq:symm_1}
\end{eqnarray}
Hence Eq.~(\ref{Eq:vel_aut_1step}) can be simplified to
\begin{equation}
                \left\langle   v_0  v_n \right\rangle =
                                                             \left( \begin{array}{cc}
                                                             1 & -1 \end{array} \right)
                                                             \left( \begin{array}{cc}
                                                             P_{11} &  P_{1-1} \\
                                                             P_{1-1} &  P_{11} \end{array} \right)^n\
                                                             \left( \begin{array}{c}
                                                             1 \\
                                                             -1 \end{array} \right)p(1),
\label{Eq:vel_aut_quadform}
\end{equation}
which is a simple quadratic form. By diagonalisation the
expression for the $n^{th}$ velocity autocorrelation function is
obtained to
\begin{equation}
                \left\langle   v_0  v_n \right\rangle = 2p(1)\left(P_{11}-P_{1-1}\right)^n
\label{Eq:vel_1step}
\end{equation}
yielding the exponential decay of the velocity autocorrelation
function $<v_0v_n>\sim\exp(n\log(P_{11}-P_{1-1}))$ referred to above.
Substituting Eq.~(\ref{Eq:vel_1step}) into the Taylor-Green-Kubo
formula Eq.~(\ref{Eq:TGK}) by using the fact that $p(1)=h/2$ gives
\begin{eqnarray}\nonumber
                D(h) &=& \sum_{n=0}^{\infty} \left\langle v_0 v_n \right\rangle   -\frac{1}{2} \left\langle v_0^2\right\rangle \\\nonumber
                     &=&  h \left(\sum_{n=0}^{\infty} \left(P_{11}-P_{1-1}\right)^n\right) -\frac{h}{2}\\
                     &=&  \frac{h}{1-P_{11}+P_{1-1}}-\frac{h}{2}.
\label{Eq:1_step_D}
\end{eqnarray}
The relevant parameter dependent probabilities can be worked out from
the invariant density $\rho^*(x)$ of the system and are
\begin{equation}
P_{11}=
\left\{
\begin{array}{ll}
0                     &\ \ 0 \leq h < \frac{1}{3}\\
1-\frac{(1-h)}{2h}    &\ \ \frac{1}{3} \leq h < \frac{1}{2}\\
\frac{1}{2}           &\ \ \frac{1}{2} \leq h < 1 \end{array}\right.
\label{Eq:P_11}
\end{equation}
and
\begin{equation}
P_{1-1}=
\left\{
\begin{array}{ll}
0                     &\ \ 0 \leq h < \frac{1}{3}\\
0                     &\ \ \frac{1}{3} \leq h < \frac{1}{2}\\
1- \frac{1}{2h}       &\ \ \frac{1}{2} \leq h < 1 \end{array}\right. \ \ .
\label{Eq:P_1-1}
\end{equation}
Substituting Eqs.~(\ref{Eq:P_11}) and (\ref{Eq:P_1-1}) into
Eq.~(\ref{Eq:1_step_D}) we obtain a persistent one-step memory
approximation for the diffusion coefficient of the map $M_h(x)$.
Figure (\ref{Fig:Pers}) shows a plot of the final result as a function
of the control parameter in comparison to the exact diffusion
coefficient $D(h)$.

\subsection{Two step memory approximation}
\label{subsec:twostep}

We now extend the approximation to include two steps of memory, i.e.,
the behaviour of a point at the $n^{th}$ step depends on what has
happened at the $(n-1)^{th}$ and $(n-2)^{th}$ step. Let $P(c|b,a)$ be
the conditional probability that a point has velocity $c$ given that
it had velocity $b$ at the previous step and $a$ at the step before
that with $a,b,c \in \left\{0,1,-1\right\}$. For this two step
approximation, the velocity autocorrelations are given by
\begin{equation}
                \left\langle   v_0  v_n \right\rangle = \sum_{v_0, ... , v_n} v_0 v_n  p(v_0,v_1) \prod_{i=2}^n P(v_i |v_{i-1},v_{i-2}),
\label{Eq:vel_aut_2step}
\end{equation}
where $p(a,b)$ is the probability that a point takes
velocity $a$ at the first step followed by $b$. Again we proceed by the
method of \cite{Gil10} and rewrite Eq.~(\ref{Eq:vel_aut_2step}) as a
matrix equation in order to capture the combinatorics of the sum,
 \begin{equation}
                 \left\langle   v_0  v_n \right\rangle = \underline{r}\cdot \underline{\underline{A}}^n \cdot \underline{s}
\label{Eq:2s}
\end{equation}
where $\underline{r}$ evaluates $v_n$, $\underline{s}$ evaluates $v_0
p(v_0,v_1)$ and $\underline{\underline{A}}$ is the $9 \times 9$ probability
transition matrix for the system. Unfortunately, Eq.~(\ref{Eq:2s})
cannot be evaluated analytically (see the Appendix), so
we resort to numerical evaluations. The result is depicted in
Fig.~\ref{Fig:Pers}.  We see that this method picks out the same
topological changes in the map dynamics that the previous method did
and interpolates between them, however, the convergence at these
points is not as accurate.

The strength of this method is in modeling the exponential decay of
correlations that is often found in diffusive systems, particularly in
Hamiltonian particle billiards \cite{Bal08}. When applied to these
systems the method is not restricted by dimension making it very
useful in this setting. However, generating by default an exponential
decay of correlations is not an ideal approach for diffusive systems
in which correlations do not decay exponentially. In contrast to the
correlated random walk approach, this method is not designed to reveal
possibly fractal structures of parameter dependent diffusion
coefficients. It also requires a lot of input about the relevant
transition probabilities, making it unpractical when it comes to
analysing higher order approximations. In particular, if one was
to consider nonhyperbolic systems \cite{KoKl03}, or even hyperbolic
systems with less simple invariant measures
\cite{RKD,RKdiss,KlDo99,KlKo02}, then deriving the transition
probabilities of Eqs.~(\ref{Eq:P_11}) and (\ref{Eq:P_1-1}) would be
much more complicated.

\section{Approximating Markov partitions}
\label{sec:approx_mark}

The final method that we look at does not involve the
Taylor-Green-Kubo formula. Using the framework of the escape rate
theory applied to dynamical systems
\cite{GN,GaDo95,BTV96,Gasp,Do99,Voll02}, we consider a truncated map
$M_h(x)$ defined on $[0,L]$. By applying absorbing boundary
conditions to this map, thus generating an open system, standard
escape rate theory expresses the diffusion coefficient in terms of the
escape rate from a fractal repeller \cite{KlDo99}. Here we use a
slight variation of this approach by using periodic boundaries. For
calculating diffusion coefficients in simple maps this setting is
technically easier, because it produces simpler transfer operators
than absorbing boundaries \cite{RKdiss}. We thus consider a closed
system, whose initial density decays exponentially to an invariant
one, as quantified by the parameter-dependent decay rate
$\gamma_{dec}(h)$.  As was shown in \cite{RKdiss,RKD,KlDo99}, by this
modified approach, and in complete analogy to ordinary escape rate
theory, the diffusion coefficient can be obtained to

\begin{equation}
               D(h) = \lim_{L\rightarrow \infty} \frac{L^2}{4\pi^2}\gamma_{dec}(h).
\label{Eq:D_L}
\end{equation}
The decay rate can in turn be calculated exactly if the
Frobenius-Perron equation can be mapped onto a Markov transition
matrix. In case of $M_h(x)$ the second largest eigenvalue
$\chi_1(h)$ of this transition matrix determines the decay rate
according to \cite{RKdiss,RKD,KlDo99}
\begin{equation}
               \gamma_{dec}(h) = ln\left(  \frac{2}{\chi_1(h)}    \right).
\label{Eq:gamma}
\end{equation}
Unfortunately, constructing Markov transition matrices exactly for
even the simplest parameter dependent maps can be a very complicated
task. Our approximate method starts as follows (see
\cite{RKdiss,KlDo99} for details): For a given value of the parameter
$h$, we restrict the dynamics to the unit interval by using
Eq.~(\ref{Eq:M_h_box}) modulo $1$. We then consider the set of
iterates of the critical point $x=0.5$, which for certain parameter
values form a set of Markov partition points. This set is then copied
and lifted back onto the system of size $L$ into each unit
interval. By supplementing this partition with periodic boundary
conditions, it defines a Markov partition for the whole system on
$[0,L]$. The key problem is that the behaviour of the orbit of the
critical point under parameter variation is very irregular. Therefore
we approximate Markov partitions by truncating this orbit for a given
parameter value after a certain number of iterations. Typically, the
resulting set of points will then not yield a Markov partition for
this parameter value. In order to make up for this, we introduce a
weighted approximation into our transition matrix to account for any
non-Markovian behaviour. For example, if partition part $i$ gets
mapped onto a fraction of partition part $j$ then the entry $a_{i,j}$
in the approximate transition matrix will be equal to this fraction;
see also \cite{Kla06} for the basic idea of this approach.

The motivation behind this method is that at each stage of the
approximation, whose level is defined by the number of iterates of the
critical point, there will be certain values of the parameter whose
Markov partitions are exact. So at least for these parameter values we
will obtain the precise diffusion coefficient $D(h)$, with
interpolations between these points as defined by the approximate
transition matrix. That way, we will have full control and
understanding over the convergence of our approximations.

\begin{figure}[t]
\begin{center}
\includegraphics[width=6cm,height=6cm]{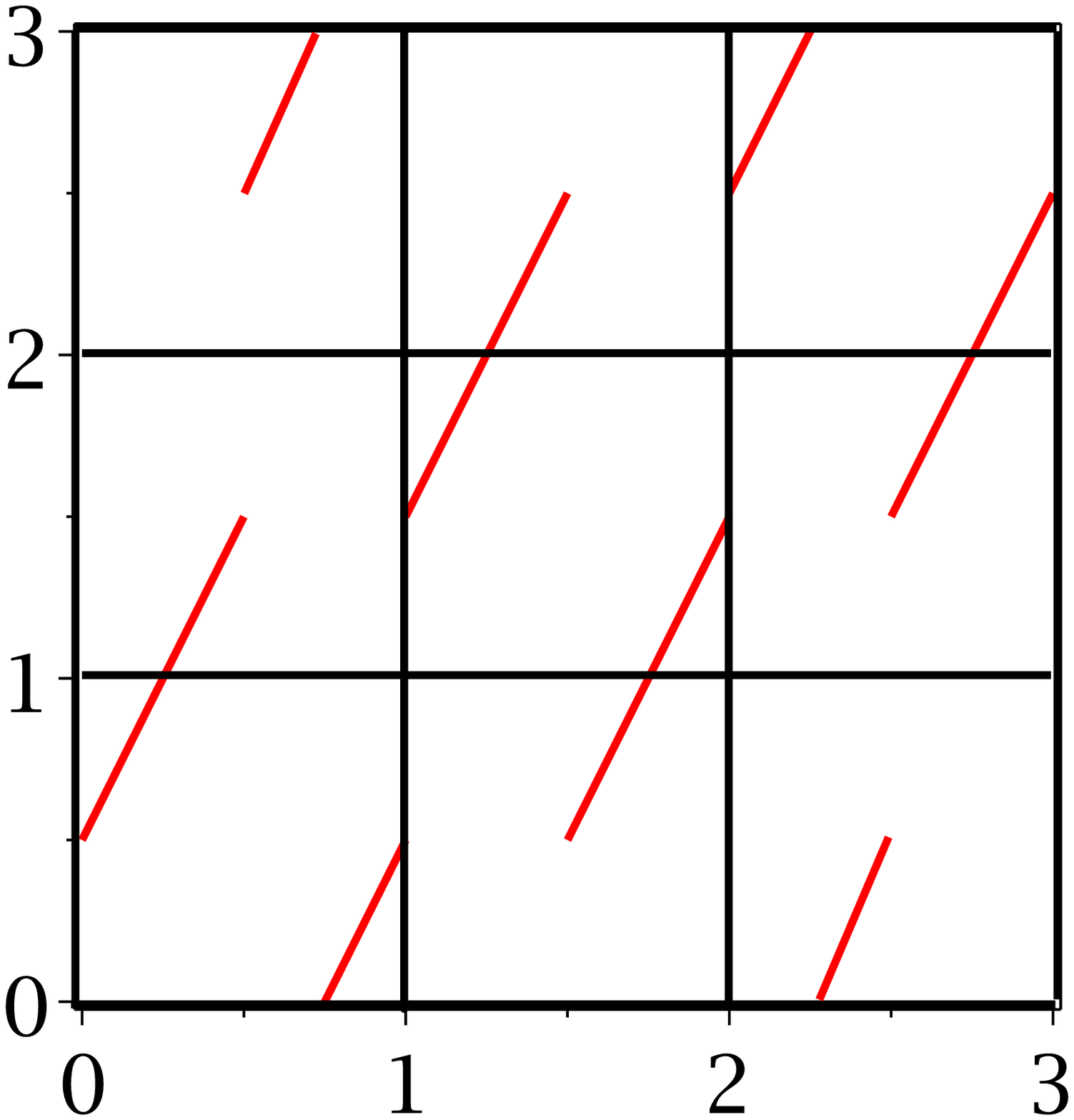}
\begin{equation}\nonumber
  \hspace*{-0.9cm}              \underline{\underline{T}}(h)=\left( \begin{array}{ccc}
                2-2h &  h    &  h \\
                h    &  2-2h &  h \\
                h    &  h    &  2-2h \end{array} \right)\
\end{equation}
\end{center}
\caption{(Colour online) Approximate Markov transition matrix . Illustrated here is the
map $M_h(x)$ truncated on $[0,L]$ with $L=3$ and periodic boundary
conditions. The map is given by the diagonal lines (red) and the zeroth order approximation
to the Markov partition is shown by the thick black lines. The
partition parts are simply the unit intervals. Note the periodic
boundary conditions. The corresponding transition matrix $\underline{\underline{T}}(h)$ is
shown below the map. Note that this partition is only Markov when $h=0$ or
$1$.}
\label{Fig:L3}
\end{figure}

We first work out the zeroth order approximation, for which we take
the unit intervals as partition parts; see Fig.~\ref{Fig:L3} for an
illustration of $M_h(x)$ at system size $L=3$. The corresponding
approximate transition matrix $\underline{\underline{T}}(h)$ is cyclic and reads

\begin{equation}
\underline{\underline{T}}(h)=\left( \begin{array}{ccccc}
2-2h  &  h     &  0     & \ldots& h \\
h     &  2-2h  &  h     & \ldots& 0     \\
0     &  h     &  2-2h  & h     & \ldots     \\
\vdots&  \vdots&  h     & \ddots& h     \\
h     & 0      &  \ldots& h     & 2-2h\end{array} \right) \:,
\label{Eq:Matrix}
\end{equation}
therefore the eigenvalues can be evaluated analytically \cite{RKD,KlDo99}
as
\begin{eqnarray}\nonumber
               \chi_1(h)&=& 2-2h +2h\cos(2\pi/L)\\
&\simeq& 2-2h+ 2h\left(1-\frac{2\pi^2}{L^2}\right)\:  (L\rightarrow\infty)\:.
\label{Eq:Eigen}
\end{eqnarray}
By combining this result with Eq.~(\ref{Eq:gamma}), the decay rate is
given as a function of the parameter and length $L$ to
\begin{equation}
              \gamma_{dec}(h)= \ln\left(\frac{1}{1-h+h\cos(2\pi/L)}\right)\simeq \frac{h2\pi^2}{L^2}\: (L\rightarrow\infty).
\label{Eq:dec}
\end{equation}
Using Eq.~(\ref{Eq:dec}) in Eq.~(\ref{Eq:D_L}), the diffusion
coefficient is finally given by
\begin{equation}
                 D(h)=\frac{h}{2}\:,
\label{Eq:mark_diff}
\end{equation}
which again yields the familiar random walk approximation.

\begin{figure*}[htb]
\begin{center}
  \includegraphics[height=6cm]{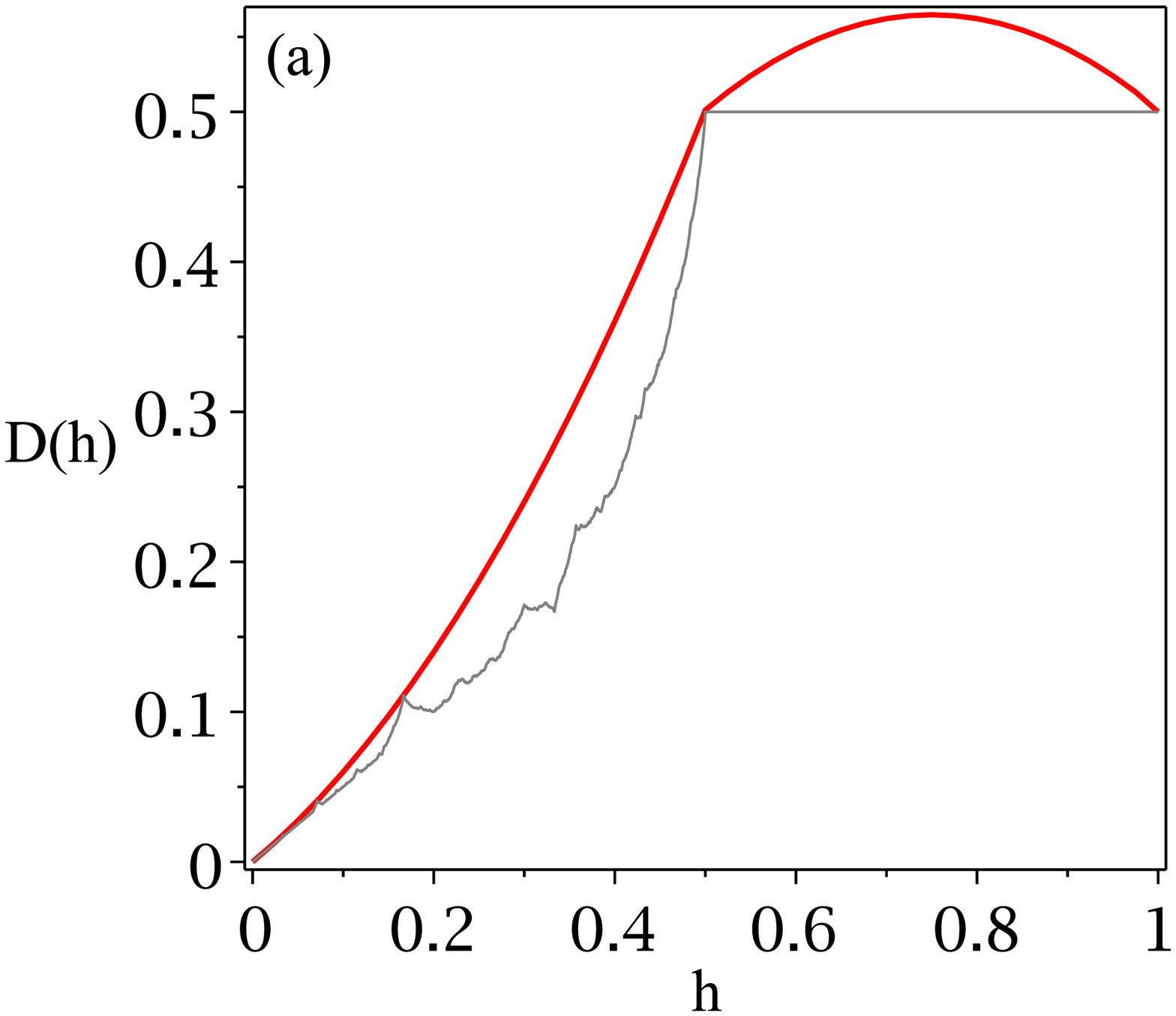}  \includegraphics[height=6cm]{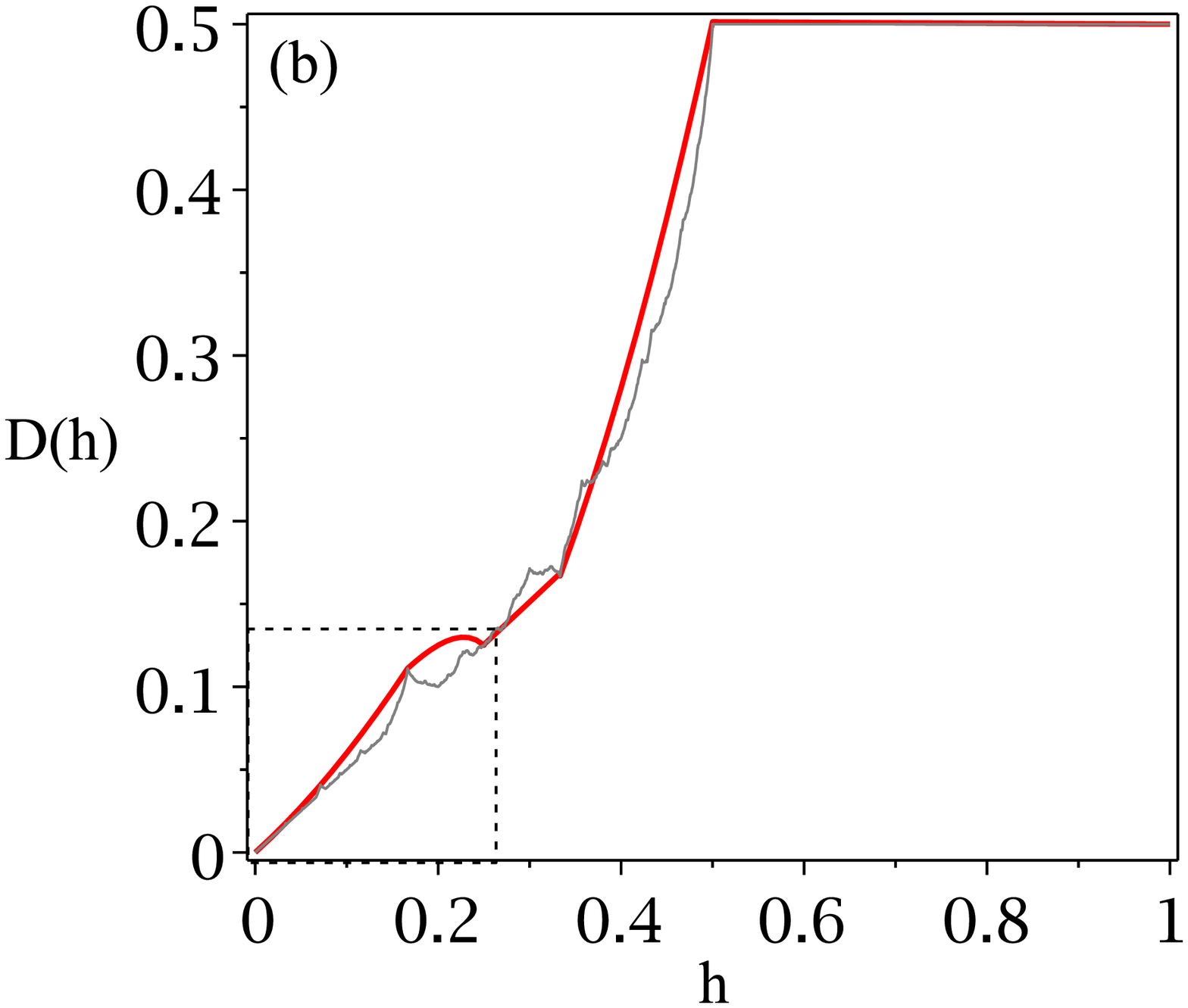} \\
  \includegraphics[height=6cm]{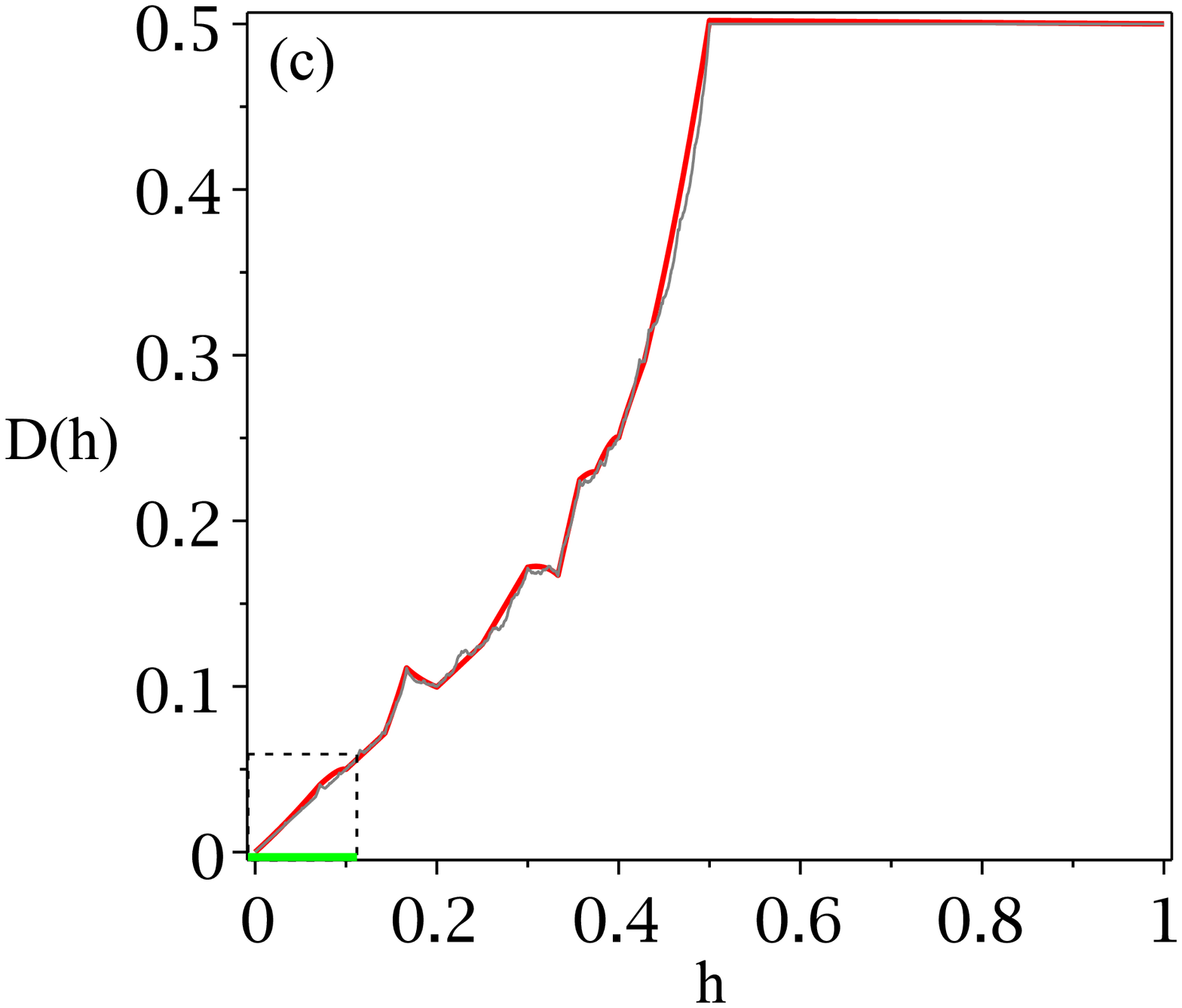}  \includegraphics[height=6cm]{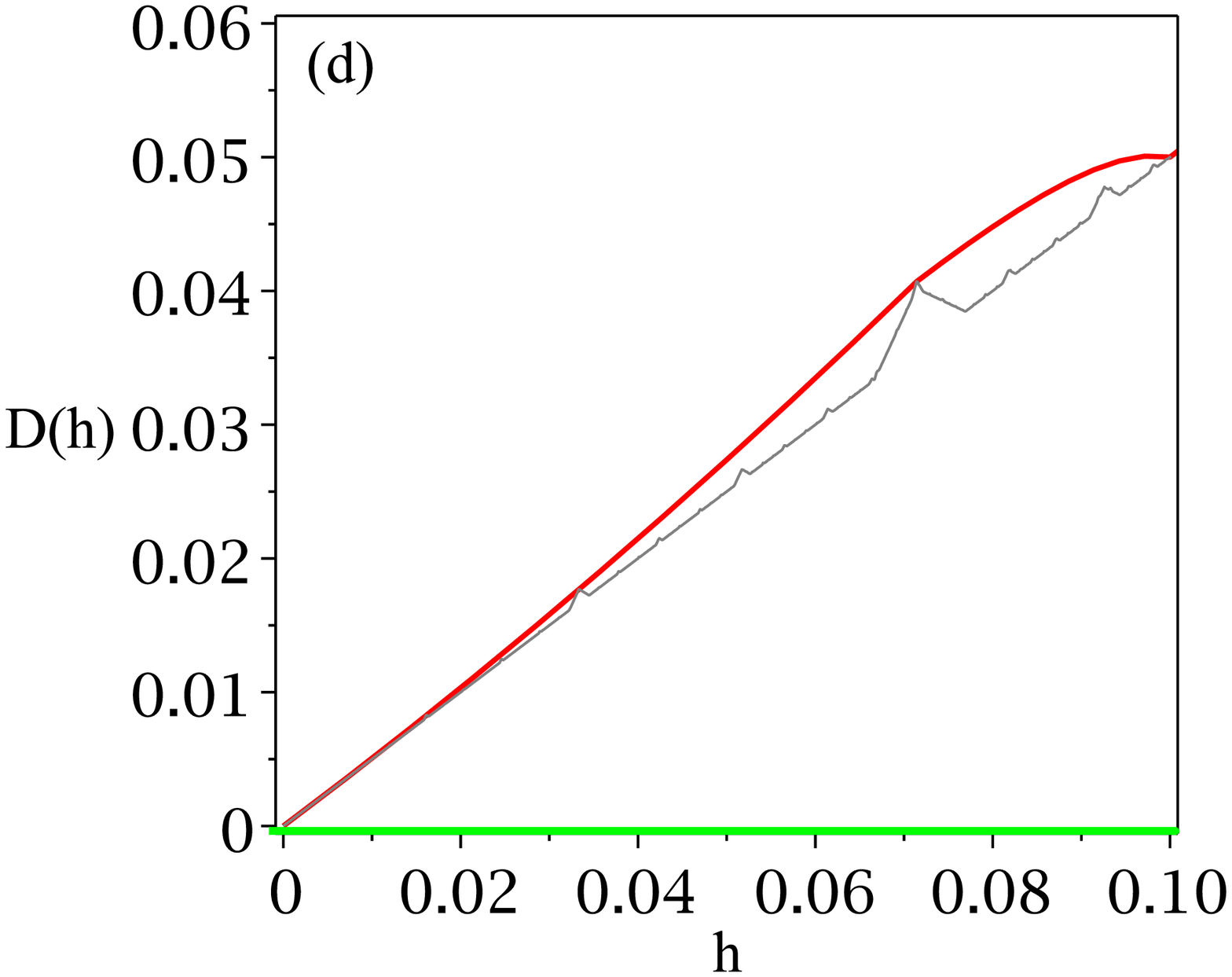}
\end{center}
\caption{(Colour online) Approximating transition matrices . In this figure, the first
order approximation to the parameter dependent diffusion coefficient
$D(h)$ obtained by this method is illustrated in \textbf{(a)}, and the
second and third orders are illustrated in \textbf{(b)} and
\textbf{(c)}, respectively, whilst a blow up of \textbf{(c)} is shown
in \textbf{(d)}. The approximations are shown in bold (red) along with the diffusion coefficient
diffusion coefficient. We see that the functional form of the
interpolation in \textbf{(a)} is repeated in \textbf{(b)} at a smaller
scale (see the contents of the dashed line box). This functional form
is again repeated on a still smaller scale in \textbf{(c)} as
illustrated in \textbf{(d)}. This self-similarity provides evidence
that the final function $D(h)$ is fractal.}
\label{Fig:Trans_mat}
\end{figure*}

The next stage of approximation involves two partition parts per unit
interval, and for this we simply include the critical point $x=0.5$ as
a partition point. So our partition parts are the half-unit intervals
on the real line. For the next iteration level we include the first
iteration of $x=0.5$, $\tilde{M}_h(0.5)=1-h$ as a partition point and
its mirror image about $x=0.5$ which is $h$, and for each higher
approximations we include one more iterate. However, with these higher
approximations we no longer obtain a cyclic matrix, so we have to
resort to numerics to evaluate the decay rate and the diffusion
coefficient. The first three approximations obtained by this method
are displayed in Fig.~\ref{Fig:Trans_mat}.

The main strength of this method is that we know by definition where
our approximations are going to converge exactly in finite time,
namely at Markov partition parameter values $h$ picked out by each
subsequent approximation. In addition, the functional form of the
interpolation between these points highlights areas of self-similarity
and therefore gives one evidence for fractal behaviour even at
low-level approximations, see Fig.~\ref{Fig:Trans_mat}. However, this
method quickly relies on numerical computation and again requires
considerable input from the user making it unpractical at higher level
approximations. It seems unlikely that this method can easily be
generalized to higher-dimensional systems, due to the difficulty to
construct (approximate) Markov partitions and associated transition
matrices.

\section{Conclusion}
\label{sec:conclusion}

We have studied three different approximation methods applied to a
particular chaotic dynamical system. For this model the exact
parameter-dependent diffusion coefficient was known beforehand
\cite{Kni11}. Taking it as a reference, the motivation was to learn
about the capabilities and the weaknesses of the individual
methods. These are of course not a comprehensive list of the many
possible ways to approximate the diffusion coefficient of a system,
see, for example, \cite{ReWi80,Ven07} for diffusion in sawtooth and
standard maps. However, what they do illustrate is the fact that even
in our simple one-dimensional model studied here, the results that one
obtains are very much dependent upon the individual method that one
uses, and these results vary greatly between the different methods.

By the first method, which relied on a systematic truncation of the
Taylor-Green-Kubo formula, we saw the fractal structure building up
fully analytically over a series of correlated random walk
approximations, as we were able to exactly capture the correlations of
the system in finite time at certain parameter values. This yielded in
turn quick convergence to the exact results. Using a persistent random
walk approach, the second method retained an exponential decay of
correlations even in finite time approximations. However, for the
model under consideration this approximation yielded convergence that
was significantly weaker than in case of the other two methods.
By using a variation of the escape rate approach to chaotic diffusion
combined with approximate transition matrices, the third method had
our attention focused on areas of self-similarity giving us
particularly strong evidence for fractal structures in the parameter
dependent diffusion coefficient. This method generated again very
quick convergence. Comparing the three different methods with each
other demonstrates that one is able to tailor the approximate results
one gets by applying a specific method to the specific questions one
wishes to answer, or to the specific setting.

As a side aspect, we addressed recent criticism of the first method by
Gilbert and Sanders \cite{Gil09}. Here we chose a different class of
systems than the Hamiltonian particle billiards that they considered
in their paper. This had the advantage that the different
approximation methods could be studied more rigorously. We conclude
that the persistent random walk method favoured in \cite{Gil09} may be
more appropriate for dispersing billiards, because an integral part of
this method is modeling an exponential decay of correlations, as it is
quite common in these systems. However, one may question the
usefulness of this method for diffusive dynamical systems where
exponential decay is not guaranteed. Here other methods, such as the
first and the third one discussed above, may yield superior results in
terms of speed of convergence and identification of possible fractal
structures in diffusion coefficients. Particularly the first method
has the advantage that it is conceptually very simple and quite
universally applicable, without making any assumptions on the decay of
correlations.

Accordingly, we find the quest for a `unique' way to approximate the
diffusion coefficient of a dynamical system, as suggested in
\cite{Gil09}, unnecessarily restrictive. In our view, each
of the three approximation methods discussed here has, for a given
model, its own virtue. When one looks to understand, or display, a
particular property of a system and cannot achieve this analytically,
resorting to one of these approximation methods is thus a sensible
course of action.

We finally emphasize that the structure of the diffusion coefficients
in more physical systems such as Lorentz gases and sawtooth maps are
still not fully understood
\cite{Kla06,Ven07}. Particularly, to which extent these systems'
diffusion coefficients are fractal remains an open question. Further
refining approximation methods, such as the ones presented in this
paper, to highlight areas of self-similarity in parameter dependent
diffusion coefficients in these systems, or to show the emergence of
fractal structures, would be of great help in answering these
questions.

\appendix

\section{Recurrence relation for the two-step approximation}

Equation (\ref{Eq:2s}) in a more explicit form is written as

\begin{widetext}
\begin{equation}\nonumber
               \left\langle   v_0  v_n \right\rangle = \left( \begin{array}{c}
                                                             0 \\
                                                             0 \\
                                                             0 \\
                                                             1 \\
                                                             1 \\
                                                             1 \\
                                                             -1\\
                                                             -1\\
                                                             -1\\ \end{array} \right)^T
                                                             \left( \begin{array}{ccccccccc}
                                                             P_{000} &  P_{001} &  P_{00-1} & 0      &  0       &  0        &  0       &  0        &  0\\
                                                             0       &  0       &  0        &P_{010} &  P_{011} &  P_{01-1} &  0       &  0        &  0\\
                                                             0       &  0       &  0        &0       &  0       &  0        & P_{0-10} &  P_{0-11} &  P_{0-1-1}\\
                                                             P_{100} &  P_{101} &  P_{10-1} & 0      &  0       &  0        &  0       &  0        &  0\\
                                                             0       &  0       &  0        &P_{110} &  P_{111} &  P_{11-1} &  0       &  0        &  0\\
                                                             0       &  0       &  0        &0       &  0       &  0        & P_{1-10} &  P_{1-11} &  P_{1-1-1}\\
                                                             P_{-100} &  P_{-101} &  P_{-10-1} & 0      &  0       &  0        &  0       &  0        &  0\\
                                                             0       &  0       &  0        &P_{-110} &  P_{-111} &  P_{-11-1} &  0       &  0        &  0\\
                                                             0       &  0       &  0        &0       &  0       &  0        & P_{-1-10} &  P_{-1-11} &  P_{-1-1-1}
                                                             \end{array} \right)^n\
                                                             \left( \begin{array}{c}
                                                             0 \\
                                                             p(1,0) \\
                                                             -p(-1,0)\\
                                                              0 \\
                                                             p(1,1) \\
                                                             -p(-1,1)\\
                                                              0 \\
                                                             p(1,-1) \\
                                                             -p(-1,-1)
                                                              \end{array} \right),
\label{Eq:2s_full}
\end{equation}
\end{widetext}
which can be simplified using the symmetries in the probabilities like
$P_{100}=P_{-100}$ and $p(1,0)=p(-1,0)$. In addition, if we let
$m_{ij}^n$ be the $ij^{th}$ entry of $\underline{\underline{A}}^n$ we
can use the fact that the symmetries of $\underline{\underline{A}}$
are the same as $\underline{\underline{A}}^n$ and reduce Eq.~
(\ref{Eq:2s}) to a $3$x$3$ matrix equation,
\begin{widetext}
\begin{equation}
                \left\langle   v_0  v_n \right\rangle =
                                                             \left( \begin{array}{ccc}
                                                             2 & 2 & 2 \end{array} \right)
                                                             \left( \begin{array}{ccc}
                                                             (m_{42}^n-m_{43}^n) &  (m_{45}^n-m_{49}^n) &  (m_{48}^n-m_{46}^n) \\
                                                             (m_{52}^n-m_{53}^n) &  (m_{55}^n-m_{59}^n) &  (m_{58}^n-m_{56}^n) \\
                                                             (m_{62}^n-m_{63}^n) &  (m_{65}^n-m_{69}^n) &  (m_{68}^n-m_{66}^n) \end{array} \right)\
                                                             \left( \begin{array}{c}
                                                             p(1,0) \\
                                                             p(1,1) \\
                                                             p(1,-1) \end{array} \right).
\label{Eq:vel_aut_mat2}
\end{equation}
\end{widetext}
In order to obtain an analytical expression for the matrix
in Eq.~(\ref{Eq:vel_aut_mat2}), we
would like to obtain a solvable recurrence relation, however, this
matrix is equal to
\begin{widetext}
\begin{eqnarray}\nonumber
                && \left( \begin{array}{ccc}
                P_{100} &  P_{101} &  P_{10-1} \\
                0 &  0 &  0 \\
                0 &  0 &  0 \end{array} \right)
                \left( \begin{array}{ccc}
                0 &  0 &  0 \\
                (m_{33}^{n-1}-m_{32}^{n-1}) &  (m_{39}^{n-1}-m_{35}^{n-1}) &  (m_{36}^{n-1}-m_{38}^{n-1}) \\
                (m_{32}^{n-1}-m_{33}^{n-1}) &  (m_{35}^{n-1}-m_{39}^{n-1})&  (m_{38}^{n-1}-m_{36}^{n-1}) \end{array} \right)\\\nonumber
                &+&\left( \begin{array}{ccc}
                0 &  0 &  0 \\
                P_{110} &  P_{111} &  P_{-1-11} \\
                0 &  0 &  0 \end{array} \right)
                \left( \begin{array}{ccc}
                (m_{42}^{n-1}-m_{43}^{n-1}) &  (m_{45}^{n-1}-m_{49}^{n-1}) &  (m_{48}^{n-1}-m_{46}^{n-1}) \\
                (m_{52}^{n-1}-m_{53}^{n-1}) &  (m_{55}^{n-1}-m_{59}^{n-1}) &  (m_{58}^{n-1}-m_{56}^{n-1}) \\
                (m_{62}^{n-1}-m_{63}^{n-1}) &  (m_{65}^{n-1}-m_{69}^{n-1})&  (m_{68}^{n-1}-m_{66}^{n-1}) \end{array} \right)\\\nonumber
                &+&\left( \begin{array}{ccc}
                0 &  0 &  0 \\
                0 &  0 &  0 \\
                P_{1-10} &  P_{1-11} &  P_{1-1-1} \end{array} \right)
                \left( \begin{array}{ccc}
                (m_{43}^{n-1}-m_{42}^{n-1}) &  (m_{49}^{n-1}-m_{45}^{n-1}) &  (m_{46}^{n-1}-m_{48}^{n-1})\\
                (m_{53}^{n-1}-m_{62}^{n-1}) &  (m_{69}^{n-1}-m_{65}^{n-1}) &  (m_{66}^{n-1}-m_{68}^{n-1}) \\
                (m_{63}^{n-1}-m_{52}^{n-1}) &  (m_{59}^{n-1}-m_{55}^{n-1})&  (m_{56}^{n-1}-m_{58}^{n-1}) \end{array} \right)
\label{Eq:2sma_rec}
\end{eqnarray}
\end{widetext}
and unlike for the one-step approximation, a recurrence relation is
unobtainable, which is due to the introduction of a zero state in
the velocities.


\begin{thebibliography}{38}
\expandafter\ifx\csname natexlab\endcsname\relax\def\natexlab#1{#1}\fi
\expandafter\ifx\csname bibnamefont\endcsname\relax
  \def\bibnamefont#1{#1}\fi
\expandafter\ifx\csname bibfnamefont\endcsname\relax
  \def\bibfnamefont#1{#1}\fi
\expandafter\ifx\csname citenamefont\endcsname\relax
  \def\citenamefont#1{#1}\fi
\expandafter\ifx\csname url\endcsname\relax
  \def\url#1{\texttt{#1}}\fi
\expandafter\ifx\csname urlprefix\endcsname\relax\def\urlprefix{URL }\fi
\providecommand{\bibinfo}[2]{#2}
\providecommand{\eprint}[2][]{\url{#2}}

\bibitem[{\citenamefont{Dorfman}(1999)}]{Do99}
\bibinfo{author}{\bibfnamefont{J.}~\bibnamefont{Dorfman}},
  \emph{\bibinfo{title}{An introduction to chaos in nonequilibrium statistical
  mechanics}} (\bibinfo{publisher}{Cambridge University Press},
  \bibinfo{address}{Cambridge}, \bibinfo{year}{1999}).

\bibitem[{\citenamefont{Gaspard}(1998)}]{Gasp}
\bibinfo{author}{\bibfnamefont{P.}~\bibnamefont{Gaspard}},
  \emph{\bibinfo{title}{Chaos, scattering, and statistical mechanics}}
  (\bibinfo{publisher}{Cambridge University Press},
  \bibinfo{address}{Cambridge}, \bibinfo{year}{1998}).

\bibitem[{\citenamefont{Klages}(2007)}]{Kla06}
\bibinfo{author}{\bibfnamefont{R.}~\bibnamefont{Klages}},
  \emph{\bibinfo{title}{Microscopic chaos, fractals and transport in
  nonequilibrium statistical mechanics}}, vol.~\bibinfo{volume}{24} of
  \emph{\bibinfo{series}{Advanced Series in Nonlinear Dynamics}}
  (\bibinfo{publisher}{World Scientific}, \bibinfo{address}{Singapore},
  \bibinfo{year}{2007}).

\bibitem[{\citenamefont{Cvitanovi{\'c}
  et~al.}(2009)\citenamefont{Cvitanovi{\'c}, Artuso, Mainieri, Tanner, and
  Vattay}}]{CAMTV01}
\bibinfo{author}{\bibfnamefont{P.}~\bibnamefont{Cvitanovi{\'c}}},
  \bibinfo{author}{\bibfnamefont{R.}~\bibnamefont{Artuso}},
  \bibinfo{author}{\bibfnamefont{R.}~\bibnamefont{Mainieri}},
  \bibinfo{author}{\bibfnamefont{G.}~\bibnamefont{Tanner}}, \bibnamefont{and}
  \bibinfo{author}{\bibfnamefont{G.}~\bibnamefont{Vattay}},
  \emph{\bibinfo{title}{Chaos: Classical and quantum}}
  (\bibinfo{publisher}{Niels Bohr Institute}, \bibinfo{address}{Copenhagen},
  \bibinfo{year}{2009}), \urlprefix\url{http://chaosbook.org}.

\bibitem[{\citenamefont{Geisel and Nierwetberg}(1982)}]{GeNi82}
\bibinfo{author}{\bibfnamefont{T.}~\bibnamefont{Geisel}} \bibnamefont{and}
  \bibinfo{author}{\bibfnamefont{J.}~\bibnamefont{Nierwetberg}},
  \bibinfo{journal}{Phys. Rev. Lett.} \textbf{\bibinfo{volume}{48}},
  \bibinfo{pages}{7} (\bibinfo{year}{1982}).

\bibitem[{\citenamefont{Schell et~al.}(1982)\citenamefont{Schell, Fraser, and
  Kapral}}]{SFK}
\bibinfo{author}{\bibfnamefont{M.}~\bibnamefont{Schell}},
  \bibinfo{author}{\bibfnamefont{S.}~\bibnamefont{Fraser}}, \bibnamefont{and}
  \bibinfo{author}{\bibfnamefont{R.}~\bibnamefont{Kapral}},
  \bibinfo{journal}{Phys. Rev. A} \textbf{\bibinfo{volume}{26}},
  \bibinfo{pages}{504} (\bibinfo{year}{1982}).

\bibitem[{\citenamefont{Fujisaka and Grossmann}(1982)}]{GF2}
\bibinfo{author}{\bibfnamefont{H.}~\bibnamefont{Fujisaka}} \bibnamefont{and}
  \bibinfo{author}{\bibfnamefont{S.}~\bibnamefont{Grossmann}},
  \bibinfo{journal}{Z. Physik B} \textbf{\bibinfo{volume}{48}},
  \bibinfo{pages}{261} (\bibinfo{year}{1982}).

\bibitem[{\citenamefont{Korabel and Klages}(2004)}]{KoKl03}
\bibinfo{author}{\bibfnamefont{N.}~\bibnamefont{Korabel}} \bibnamefont{and}
  \bibinfo{author}{\bibfnamefont{R.}~\bibnamefont{Klages}},
  \bibinfo{journal}{Physica D} \textbf{\bibinfo{volume}{187}},
  \bibinfo{pages}{66} (\bibinfo{year}{2004}).

\bibitem[{\citenamefont{Rechester and White}(1980)}]{ReWi80}
\bibinfo{author}{\bibfnamefont{A.B.}~\bibnamefont{Rechester}} \bibnamefont{and}
  \bibinfo{author}{\bibfnamefont{R.B}~\bibnamefont{White}},
  \bibinfo{journal}{Phys. Rev. Lett.} \textbf{\bibinfo{volume}{44}},
  \bibinfo{pages}{1586} (\bibinfo{year}{1980}).

\bibitem[{\citenamefont{Cary and Meiss}(1981)}]{CM81}
\bibinfo{author}{\bibfnamefont{J.R.}~\bibnamefont{Cary}} \bibnamefont{and}
  \bibinfo{author}{\bibfnamefont{J.D.}~\bibnamefont{Meiss}},
  \bibinfo{journal}{Phys. Rev. A} \textbf{\bibinfo{volume}{24}},
  \bibinfo{pages}{2664} (\bibinfo{year}{1981}).

\bibitem[{\citenamefont{Venegeroles}(2007)}]{Ven07}
\bibinfo{author}{\bibfnamefont{R.}~\bibnamefont{Venegeroles}},
  \bibinfo{journal}{Phys. Rev. Lett.} \textbf{\bibinfo{volume}{99}},
  \bibinfo{pages}{014101} (\bibinfo{year}{2007}).

\bibitem[{\citenamefont{Machta and Zwanzig}(1983)}]{MaZw83}
\bibinfo{author}{\bibfnamefont{J.}~\bibnamefont{Machta}} \bibnamefont{and}
  \bibinfo{author}{\bibfnamefont{R.}~\bibnamefont{Zwanzig}},
  \bibinfo{journal}{Phys. Rev. Lett.} \textbf{\bibinfo{volume}{50}},
  \bibinfo{pages}{1959} (\bibinfo{year}{1983}).

\bibitem[{\citenamefont{Harayama and Gaspard}(2001)}]{HaGa01}
\bibinfo{author}{\bibfnamefont{T.}~\bibnamefont{Harayama}} \bibnamefont{and}
  \bibinfo{author}{\bibfnamefont{P.}~\bibnamefont{Gaspard}},
  \bibinfo{journal}{Phys. Rev. E} \textbf{\bibinfo{volume}{64}},
  \bibinfo{pages}{036215} (\bibinfo{year}{2001}).

\bibitem[{\citenamefont{Harayama et~al.}(2002)\citenamefont{Harayama, Klages,
  and Gaspard}}]{HaKlGa02}
\bibinfo{author}{\bibfnamefont{T.}~\bibnamefont{Harayama}},
  \bibinfo{author}{\bibfnamefont{R.}~\bibnamefont{Klages}}, \bibnamefont{and}
  \bibinfo{author}{\bibfnamefont{P.}~\bibnamefont{Gaspard}},
  \bibinfo{journal}{Phys. Rev. E} \textbf{\bibinfo{volume}{66}},
  \bibinfo{pages}{026211} (\bibinfo{year}{2002}).

\bibitem[{\citenamefont{M{\'a}ty{\'a}s and Klages}(2004)}]{MaKl03}
\bibinfo{author}{\bibfnamefont{L.}~\bibnamefont{M{\'a}ty{\'a}s}}
  \bibnamefont{and} \bibinfo{author}{\bibfnamefont{R.}~\bibnamefont{Klages}},
  \bibinfo{journal}{Physica D} \textbf{\bibinfo{volume}{187}},
  \bibinfo{pages}{165} (\bibinfo{year}{2004}).

\bibitem[{\citenamefont{Klages}(1996)}]{RKdiss}
\bibinfo{author}{\bibfnamefont{R.}~\bibnamefont{Klages}},
  \emph{\bibinfo{title}{Deterministic diffusion in one-dimensional chaotic
  dynamical systems}} (\bibinfo{publisher}{Wissenschaft \& Technik-Verlag},
  \bibinfo{address}{Berlin}, \bibinfo{year}{1996}).

\bibitem[{\citenamefont{Klages and Dorfman}(1995)}]{RKD}
\bibinfo{author}{\bibfnamefont{R.}~\bibnamefont{Klages}} \bibnamefont{and}
  \bibinfo{author}{\bibfnamefont{J.R.}~\bibnamefont{Dorfman}},
  \bibinfo{journal}{Phys. Rev. Lett.} \textbf{\bibinfo{volume}{74}},
  \bibinfo{pages}{387} (\bibinfo{year}{1995}).

\bibitem[{\citenamefont{Klages and Dorfman}(1999)}]{KlDo99}
\bibinfo{author}{\bibfnamefont{R.}~\bibnamefont{Klages}} \bibnamefont{and}
  \bibinfo{author}{\bibfnamefont{J.R.}~\bibnamefont{Dorfman}},
  \bibinfo{journal}{Phys. Rev. E} \textbf{\bibinfo{volume}{59}},
  \bibinfo{pages}{5361} (\bibinfo{year}{1999}).

\bibitem[{\citenamefont{Groeneveld and Klages}(2002)}]{GrKl02}
\bibinfo{author}{\bibfnamefont{J.}~\bibnamefont{Groeneveld}} \bibnamefont{and}
  \bibinfo{author}{\bibfnamefont{R.}~\bibnamefont{Klages}},
  \bibinfo{journal}{J. Stat. Phys.} \textbf{\bibinfo{volume}{109}},
  \bibinfo{pages}{821} (\bibinfo{year}{2002}).

\bibitem[{\citenamefont{Cristadoro}(2006)}]{Crist06}
\bibinfo{author}{\bibfnamefont{G.}~\bibnamefont{Cristadoro}},
  \bibinfo{journal}{J. Phys. A: Math. Gen.} \textbf{\bibinfo{volume}{39}},
  \bibinfo{pages}{L151} (\bibinfo{year}{2006}).

\bibitem[{\citenamefont{Knight and Klages}(2011)}]{Kni11}
\bibinfo{author}{\bibfnamefont{G.}~\bibnamefont{Knight}} \bibnamefont{and}
  \bibinfo{author}{\bibfnamefont{R.}~\bibnamefont{Klages}},
  \bibinfo{journal}{Nonlinearity} \textbf{\bibinfo{volume}{24}},
  \bibinfo{pages}{227} (\bibinfo{year}{2011}).

\bibitem[{\citenamefont{{Jepps} and {Rondoni}}(2006)}]{JeRo06}
\bibinfo{author}{\bibfnamefont{O.}~\bibnamefont{{Jepps}}} \bibnamefont{and}
  \bibinfo{author}{\bibfnamefont{L.}~\bibnamefont{{Rondoni}}},
  \bibinfo{journal}{J. Phys. A: Math. Gen.} \textbf{\bibinfo{volume}{39}},
  \bibinfo{pages}{1311} (\bibinfo{year}{2006}).

\bibitem[{\citenamefont{Dana et~al.}(1989)\citenamefont{Dana, Murray, and
  Percival}}]{DMP89}
\bibinfo{author}{\bibfnamefont{I.}~\bibnamefont{Dana}},
  \bibinfo{author}{\bibfnamefont{N.W.}~\bibnamefont{Murray}}, \bibnamefont{and}
  \bibinfo{author}{\bibfnamefont{I.C}~\bibnamefont{Percival}},
  \bibinfo{journal}{Phys. Rev. Lett.} \textbf{\bibinfo{volume}{62}},
  \bibinfo{pages}{233} (\bibinfo{year}{1989}).

\bibitem[{\citenamefont{Eckhardt}(1993)}]{Eck92}
\bibinfo{author}{\bibfnamefont{B.}~\bibnamefont{Eckhardt}},
  \bibinfo{journal}{Phys. Lett. A} \textbf{\bibinfo{volume}{172}},
  \bibinfo{pages}{411 } (\bibinfo{year}{1993}).

\bibitem[{\citenamefont{Percival and Vivaldi}(1987{\natexlab{a}})}]{PV86}
\bibinfo{author}{\bibfnamefont{I.}~\bibnamefont{Percival}} \bibnamefont{and}
  \bibinfo{author}{\bibfnamefont{F.}~\bibnamefont{Vivaldi}},
  \bibinfo{journal}{Physica D} \textbf{\bibinfo{volume}{27}},
  \bibinfo{pages}{373 } (\bibinfo{year}{1987}{\natexlab{a}}).

\bibitem[{\citenamefont{Percival and Vivaldi}(1987{\natexlab{b}})}]{PV862}
\bibinfo{author}{\bibfnamefont{I.}~\bibnamefont{Percival}} \bibnamefont{and}
  \bibinfo{author}{\bibfnamefont{F.}~\bibnamefont{Vivaldi}},
  \bibinfo{journal}{Physica D} \textbf{\bibinfo{volume}{25}},
  \bibinfo{pages}{105 } (\bibinfo{year}{1987}{\natexlab{b}}).

\bibitem[{\citenamefont{Sano}(2002)}]{Sano02}
\bibinfo{author}{\bibfnamefont{M.~M.} \bibnamefont{Sano}},
  \bibinfo{journal}{Phys. Rev. E} \textbf{\bibinfo{volume}{66}},
  \bibinfo{pages}{046211} (\bibinfo{year}{2002}).

\bibitem[{\citenamefont{Gilbert and Sanders}(2009)}]{Gil09}
\bibinfo{author}{\bibfnamefont{T.}~\bibnamefont{Gilbert}} \bibnamefont{and}
  \bibinfo{author}{\bibfnamefont{D.~P.} \bibnamefont{Sanders}},
  \bibinfo{journal}{Phys. Rev. E} \textbf{\bibinfo{volume}{80}},
  \bibinfo{pages}{041121} (\bibinfo{year}{2009}).

\bibitem[{\citenamefont{Klages and Korabel}(2002)}]{KlKo02}
\bibinfo{author}{\bibfnamefont{R.}~\bibnamefont{Klages}} \bibnamefont{and}
  \bibinfo{author}{\bibfnamefont{N.}~\bibnamefont{Korabel}},
  \bibinfo{journal}{J. Phys. A: Math. Gen.} \textbf{\bibinfo{volume}{35}},
  \bibinfo{pages}{4823} (\bibinfo{year}{2002}).

\bibitem[{\citenamefont{Haus and Kehr}(1987)}]{HK87}
\bibinfo{author}{\bibfnamefont{J.}~\bibnamefont{Haus}} \bibnamefont{and}
  \bibinfo{author}{\bibfnamefont{K.~W.} \bibnamefont{Kehr}},
  \bibinfo{journal}{Phys. Rep.} \textbf{\bibinfo{volume}{150}},
  \bibinfo{pages}{264} (\bibinfo{year}{1987}).

\bibitem[{\citenamefont{Weiss}(1994)}]{Weiss94}
\bibinfo{author}{\bibfnamefont{G.}~\bibnamefont{Weiss}},
  \emph{\bibinfo{title}{Aspects and applications of the random walk}}
  (\bibinfo{publisher}{North-Holland}, \bibinfo{address}{Amsterdam},
  \bibinfo{year}{1994}).

\bibitem[{\citenamefont{Gilbert and Sanders}(2010)}]{Gil10}
\bibinfo{author}{\bibfnamefont{T.}~\bibnamefont{Gilbert}} \bibnamefont{and}
  \bibinfo{author}{\bibfnamefont{D.~P.} \bibnamefont{Sanders}},
  \bibinfo{journal}{J. Phys. A: Math. Theor.} \textbf{\bibinfo{volume}{43}},
  \bibinfo{pages}{035001} (\bibinfo{year}{2010}).

\bibitem[{\citenamefont{Gilbert et~al.}(2011)\citenamefont{Gilbert, Nguyen, and
  Sanders}}]{Gil11}
\bibinfo{author}{\bibfnamefont{T.}~\bibnamefont{Gilbert}},
  \bibinfo{author}{\bibfnamefont{H.~C.} \bibnamefont{Nguyen}},
  \bibnamefont{and} \bibinfo{author}{\bibfnamefont{D.~P.}
  \bibnamefont{Sanders}}, \bibinfo{journal}{J. Phys. A: Math. Theor.} \textbf{\bibinfo{volume}{44}}, \bibinfo{pages}{065001}
  (\bibinfo{year}{2011}).

\bibitem[{\citenamefont{Gaspard and Nicolis}(1990)}]{GN}
\bibinfo{author}{\bibfnamefont{P.}~\bibnamefont{Gaspard}} \bibnamefont{and}
  \bibinfo{author}{\bibfnamefont{G.}~\bibnamefont{Nicolis}},
  \bibinfo{journal}{Phys. Rev. Lett.} \textbf{\bibinfo{volume}{65}},
  \bibinfo{pages}{1693} (\bibinfo{year}{1990}).

\bibitem[{\citenamefont{Gaspard and Dorfman}(1995)}]{GaDo95}
\bibinfo{author}{\bibfnamefont{P.}~\bibnamefont{Gaspard}} \bibnamefont{and}
  \bibinfo{author}{\bibfnamefont{J.R.}~\bibnamefont{Dorfman}},
  \bibinfo{journal}{Phys. Rev. E} \textbf{\bibinfo{volume}{52}},
  \bibinfo{pages}{3525} (\bibinfo{year}{1995}).

\bibitem[{\citenamefont{Breymann, T\'el, and Vollmer}(1996)}]{BTV96}
\bibinfo{author}{\bibfnamefont{W.}~\bibnamefont{Breymann}},
  \bibinfo{author}{\bibfnamefont{T.}~\bibnamefont{T\'el}} \bibnamefont{and}
  \bibinfo{author}{\bibfnamefont{J.}~\bibnamefont{Vollmer}},
  \bibinfo{journal}{Phys. Rev. Lett.} \textbf{\bibinfo{volume}{77}},
  \bibinfo{pages}{2945} (\bibinfo{year}{1996}).

\bibitem[{\citenamefont{Vollmer}(2002)}]{Voll02}
\bibinfo{author}{\bibfnamefont{J.}~\bibnamefont{Vollmer}}, \bibinfo{journal}{Phys.
  Rep.} \textbf{\bibinfo{volume}{372}}, \bibinfo{pages}{131}
  (\bibinfo{year}{2002}).

\bibitem[{\citenamefont{Gaspard and Klages}(1998)}]{GaKl}
\bibinfo{author}{\bibfnamefont{P.}~\bibnamefont{Gaspard}} \bibnamefont{and}
  \bibinfo{author}{\bibfnamefont{R.}~\bibnamefont{Klages}},
  \bibinfo{journal}{Chaos} \textbf{\bibinfo{volume}{8}}, \bibinfo{pages}{409}
  (\bibinfo{year}{1998}).

\bibitem[{\citenamefont{Gaspard}(1992)}]{PG1}
\bibinfo{author}{\bibfnamefont{P.}~\bibnamefont{Gaspard}}, \bibinfo{journal}{J.
  Stat. Phys.} \textbf{\bibinfo{volume}{68}}, \bibinfo{pages}{673}
  (\bibinfo{year}{1992}).

\bibitem[{\citenamefont{B\acute{a}lint and Toth}(2008)}]{Bal08}
\bibinfo{author}{\bibfnamefont{P.}~\bibnamefont{B$\acute{a}$lint}}
  \bibnamefont{and} \bibinfo{author}{\bibfnamefont{I.}~\bibnamefont{Toth}},
  \bibinfo{journal}{Ann. Inst. Henri Poincare} \textbf{\bibinfo{volume}{9}},
  \bibinfo{pages}{1309} (\bibinfo{year}{2008}).

\end{thebibliography}

\end{document}